\documentclass[useAMS,usenatbib,aas_macros]{mn2e}

\usepackage{epsfig}
\usepackage{color}
\usepackage{amssymb}
\usepackage{ams math}
\usepackage[greek,english]{babel}

\newcommand{\Lya}{Lyman-\greektext a\latintext}
\def\yjh{Y\!JH}

\title[Unbiased UV spectral slopes of $z\approx7$ galaxies]{The unbiased measurement of UV spectral slopes in low luminosity galaxies at $\bmath{z\approx7}$}
\author[A. B. Rogers, R. J. McLure and J. S. Dunlop]{\parbox\textwidth{A. B. Rogers $^{1}$\thanks{E-mail:abr@roe.ac.uk}, R. J. McLure$^{1}$ and J. S.
Dunlop$^{1}$} \\
$^{1}$SUPA\thanks{Scottish Universities Physics Alliance}, Institute for Astronomy, University of Edinburgh, Royal Observatory, Edinburgh EH9 3HJ}
\begin{document}

\date{Accepted 2012 November 28. Received 2012 November 14; in original form 2012 September 20}

\pagerange{\pageref{firstpage}--\pageref{lastpage}} \pubyear{2012}

\maketitle

\label{firstpage}

\begin{abstract}
The Ultraviolet (UV) continuum slope $\beta$, typically observed at $z\approx7$ in \emph{Hubble Space Telescope} (\emph{HST}) WFC3/IR bands via the $J-H$ colour, is a useful indicator of the age, metallicity, and dust content of high-redshift stellar populations. Recent studies have shown that the redward evolution of $\beta$ with cosmic time from redshift 7 to 4 can be largely explained by a build up of dust. However, initial claims that faint $z\approx7$ galaxies in the Hubble Ultra Deep Field WFC3/IR imaging (HUDF09) were blue enough to require stellar populations of zero reddening, low metallicity and young ages, hitherto unseen in star-forming galaxies, have since been refuted and revised. 
Here we revisit the question of how best to measure the UV slope of $z\approx7$ galaxies through source recovery simulations, within the context of present and future ultra-deep imaging from \emph{HST}.
We consider how source detection, selection and colour measurement have each biased the measurement of $\beta$ in previous studies.
After finding a robust method for measuring $\beta$ in the simulations (via a power law fit to all the available photometry), we remeasure the UV slopes of a sample of previously published low luminosity $z\approx7$ galaxy candidates.
The mean UV slope of faint galaxies in this sample appears consistent with an intrinsic distribution of normal star-forming galaxies with $\beta\approx-2$, although properly decoding the underlying distribution will require further imaging from the ongoing HUDF12 programme.
We therefore go on to consider strategies for obtaining better constraints on the underlying distribution of UV slopes at $z\approx7$ from these new data, which will benefit particularly from the addition of imaging in a second $J$-band filter: F140W. We find that a precise and unbiased measurement of $\beta$ should then be possible.
\end{abstract}


%%%%%%%%%%%%%%%%%%%%%%%%%%%%%%%%%%%%%%%%%%%%%%%%%%%%%%%%%%%%%%%%%%%%%%%%%%%%
%%%%%%%%%%%%%%%%%%%%%%%%%%%%%%%%%%%%%%%%%%%%%%%%%%%%%%%%%%%%%%%%%%%%%%%%%%%%

\section{Introduction}
Recent observations with WFC3/IR on board the \emph{Hubble Space Telescope} (\emph{HST}) have begun to probe the hitherto unconstrained properties of galaxies at $z\approx7$. 
The efficient `wedding-cake' strategy of combining wide and deep imaging from the Cosmic Assembly Near-infrared Deep Extragalactic Legacy Survey (CANDELS) programme \citep{Grogin2011, Koekemoer2011} with ultra-deep imaging in the Hubble Ultra Deep Field \citep[HUDF; e.g.][]{Bouwens2010, Bunker2010, Finkelstein2010a, Lorenzoni2011, McLure2010, McLure2011, Oesch2010, Oesch2012, Oesch2012b, Wilkins2011a, Yan2011} is now beginning to provide large, statistical, samples of $z\approx7$ galaxies with good dynamic range in luminosity. 
The faintest of these galaxies are expected to be a key source of the ionising photons required to reionise the universe, yet constraining their contribution relies on measuring their UV colours in order to estimate the galaxies' production rate of Hydrogen ionising photons and the fraction of such photons which escape the galaxy to reionise the IGM \citep[e.g.][]{Robertson2010, Finkelstein2012}.

The UV spectrum can be parametrized by the UV continuum slope $\beta$, defined by 
$f_\lambda\propto\lambda^\beta$ \citep[e.g.][]{Meurer1999}, such that a flat spectrum with zero colour in the AB magnitude system has $\beta=-2$.
A UV slope of $\beta=-2$ is typical of a young, un-reddened, low-metallicity, star-forming galaxy at $z\approx2$ \citep[e.g.][]{Erb2010}.
%e.g. BX418, a galaxy with $\beta=-2.1$ at $z=2.3$ \citep{Erb2010}.
The current interest in the UV slopes of $z\approx7$ galaxies began when the optical ACS data in the HUDF \citep{Beckwith2006} was complimented by the WFC3/IR $Y,J,H$-band data of the HUDF09 programme (GO11563, PI: Illingworth; \citealt{Bouwens2010a}).
The first substantial catalogues of $z\approx7$ Lyman Break Galaxies (LBGs) were available following the first epoch of this programme in 2009. 
Hereafter, we refer to data taken prior to and during this first HUDF09 epoch as HUDF09E1. 
Later studies have made use of further data obtained in a second epoch, and we refer to the stack of all the WFC3/IR data from epochs 1 and 2 as HUDF09FULL.
In this paper we also refer to the WFC3 Early Release Science (ERS, \citealt{Windhorst2011}) programme, which provides shallower imaging over a wider (36.5 sq. arcmin) field than the HUDF (4.5 sq. arcmin).

In an initial foray into the measurement of $\beta$ at $z\approx~7$, \citet{Bouwens2010} found evidence for a colour-magnitude relation such that the faintest $z\approx7$ galaxies exhibited 
sufficiently blue average UV colours ($\langle \beta \rangle =-3.0\pm0.2$ at $-19\leq M_{\rm UV\!,\,AB}\leq-18$) that extremely young ages and ultra-low metallicities could be inferred.
If confirmed, it would be necessary to not only account for the rapid evolution of stellar populations from $z\approx7\rightarrow 6$, but also to conclude that the UV photon escape fraction must be sufficiently high at $z\approx7$ that nebular continuum emission does not significantly redden the observed SED.
With the same dataset -- the HUDF09E1 -- \citet{Finkelstein2010a} found similarly extreme $\beta$ values, although with a sufficiently large error that it was not necessary to invoke `exotic' stellar populations. 
In fact, they suggested that only moderately young, dust-free, stellar populations are required to reproduce the observed colours.
With improved data in the final HUDF09FULL, a revised assessment reported by \citet{Bouwens2012} retains a clear colour-magnitude trend, albeit with the faintest objects averaging only $\langle\beta\rangle=-2.68\pm0.19\pm0.28$ (biweight mean $\pm$ random $\pm$ systematic uncertainties). 
Significantly, \citet{Finkelstein2012a} also find that the full HUDF09FULL dataset provides somewhat redder colours for the faintest\footnote{In HUDF09E1, this refers to galaxies in the faintest 1~mag bin; in HUDF09FULL galaxies with $L<0.25~L^{*}$.} objects with $\langle\beta\rangle=-3.07\pm0.51$ in HUDF09E1 \citep{Finkelstein2010a}, and $\langle\beta\rangle=-2.68^{+0.39}_{-0.24}$ ($\approx-2.4$ after bias correction) in HUDF09FULL \citep{Finkelstein2012a}. 
Already it should be clear from these revised estimates and their quoted uncertainties that the typical UV spectral slope of the faintest galaxies at $z\approx7$ is not well constrained.

In response to the initial claims of the discovery of exotic stellar populations at $z\approx7$, 
\citet{Dunlop2012} demonstrated that measurement biases can result in a population of normal 
star-forming galaxies with $\langle{\beta}\rangle\approx-2$ being observed as a population of extremely blue objects, especially when average properties 
are calculated for objects close to the detection limit of the imaging. This is because, as the scatter in observed colour inevitably rises 
when the flux-density limit of the survey is approached, the methods used to select LBGs (either simple colour-colour selection, or multi-band photometric redshift 
determination) can start to preferentially exclude genuine high-redshift objects whose colours have been scattered to very red values
(treating them as likely lower-redshift interlopers). It is important 
to stress that this is not the same as saying that LBG selection is limited to extremely blue objects. In fact the commonly-used colour-colour selection criteria,
and photometric-redshift selection techniques can admit quite red LBGs, especially if the photometric-redshift technique is not confined to the most 
secure candidates. Nevertheless, because photometric scatter can result in extreme (indeed often unphysical) values of $\beta$ for individual 
objects, the reddest objects can be ``lost'' to ostensibly low photometric redshifts, while the extreme blue objects never are (with resulting implications for 
the calculation of average values of $\beta$). 

Contrary to the claim made by \citet{Bouwens2012}, \citet{Dunlop2012} focussed on the inclusion of {\it all} candidate objects with even 
a marginally-preferred high-redshift LBG solution, and showed that, with existing data, there still exists a significant blue bias ($\Delta \langle \beta \rangle \approx 
-0.5$) in the inferred value 
of $\langle \beta \rangle$ for the faintest LBGs at $z \approx 7$ (the bias is simply more extreme if only the most robust LBG candidates 
are considered; e.g. \citealt{McLure2011}). As we show later in this paper, this level of bias also applies to colour-colour selected samples \citep[e.g.][]{Bouwens2012} and 
is exacerbated by the imposition of a $J$-band flux threshold in the galaxy sample selection \citep[e.g.][]{Bouwens2010}. 

\begin{figure*}
\includegraphics[width=7in]{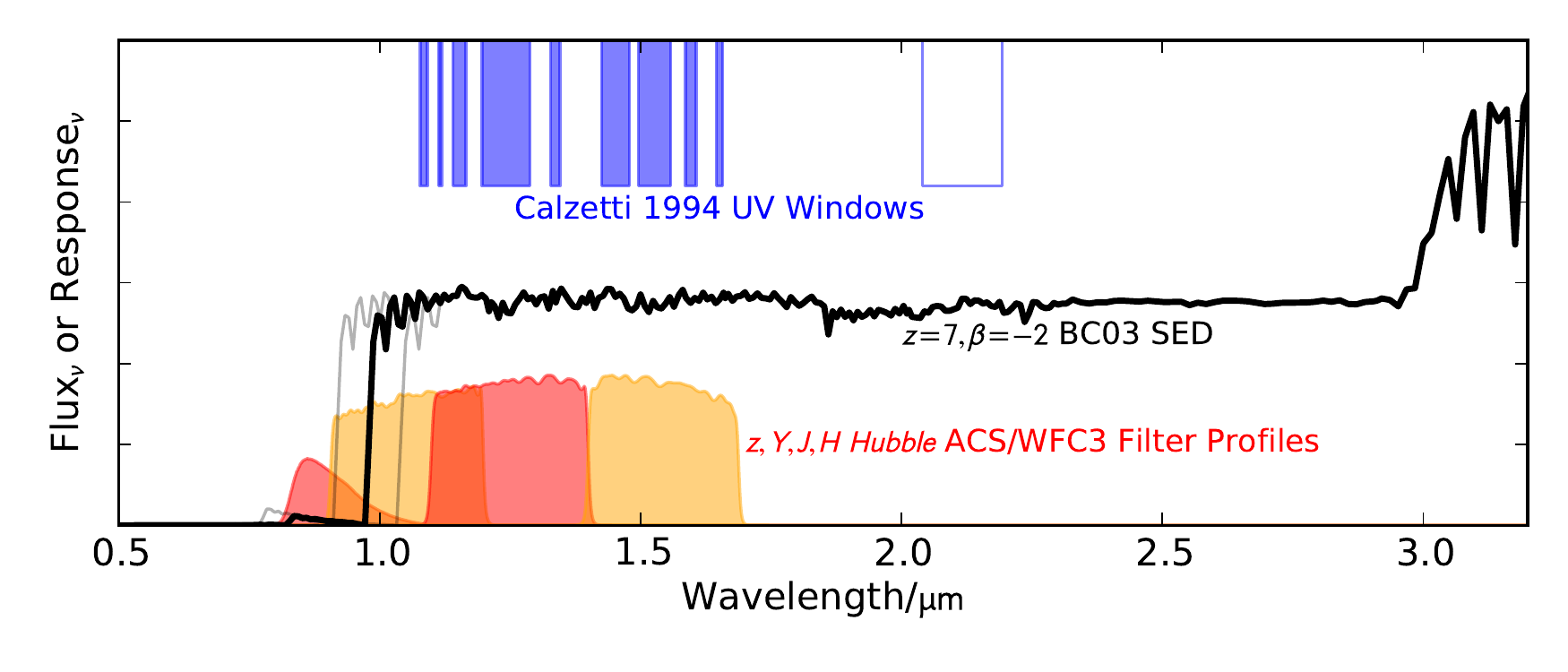}
\caption{The typical Spectral Energy Distribution (SED) of a $z=7$ Lyman Break Galaxy, and the filters used to observe it. The SED shown by a thick black line is a $\beta=-2$ \citet{Bruzual2003} 
stellar population model (in this case a 60~Myr old $0.2\,\rm{Z}_{\odot}$ star-burst), 
attenuated by \citet{Madau1995} prescribed  IGM absorption. The red and orange filter profiles shown are \emph{HST} ACS $z_{850}$, WFC3/IR $Y_{105}, J_{125}$, and $H_{160}$ -- those filters that probe the rest-frame UV at $z\approx6$ to $7$ in the HUDF09 dataset.
Shown in blue above the SED are the locations of the ten \citet{Calzetti1994} UV windows, used in the `best-fitting model' $\beta$ measurement method (see Section \ref{bestfitmodelsection}).
The unshaded \citet{Calzetti1994} window is neglected in our fitting (see text). Two grey SED curves show the Lyman break position at $z=6.5$ and $7.5$, illustrating that, within the $z\approx7$ sample, the Lyman break always attenuates the light within the $Y_{105}$-band. The $Y_{098}$ (F098M) filter used in the ERS observations approximately spans the shorter two thirds of $Y_{105}$'s wavelength coverage, without overlapping $J_{125}$.}
\label{sedandfiltersplot}
\end{figure*}

From the above discussion, it is clear that the steepness of the UV slope for sub-$L^{*}, z\approx7$ galaxies remains an open question.
The situation is further confused by the fact that different studies have used different datasets, selection methods and techniques for measuring $\beta$ (e.g. from a single near-IR colour or from SED fitting).
The first objective of this paper is therefore to use simulated data to investigate the impact of image depth, selection biases and measurement techniques on the recovered values of $\beta$.
Then, based on our findings, we explore possible strategies for the optimum analysis of the new, deeper WFC3/IR imaging
of the HUDF which will be provided by the HUDF12 project (including imaging in an additional wave-band ($J_{140}$); GO12498, PI: Ellis; public data release by early 2013), 
in order to extract the most robust, least-biased estimate of $\beta$ for the faintest LBGs at $z \approx 7$.

Throughout this paper we will consider $z\approx7$ LBGs to be objects selected with photometric redshift solutions in the range $6.5\leq z\leq7.5$.
As with previous studies in this area, we do not consider $z\approx8$ galaxies given the lack of data redward of the Lyman break\footnote{\citet{Finkelstein2012a} do provide estimates of $\beta$ at $z\approx8$, but due to the  uncertainties do not draw any conclusions.} (which begins to attenuate the $J_{125}$-band flux at $z\gtrsim7.9$).

This paper is laid out as follows. 
In Section \ref{measuringbeta}, we summarise three methods of measuring $\beta$ -- from a single colour, a power-law or the best-fitting galaxy-model SED. 
In Section \ref{simulations}, we provide a description of our simulation pipeline.
In Section \ref{comparedunlopsection} we compare with and endorse the conclusions of \citet{Dunlop2012}. 
In Section \ref{simulationresults}, we show the results of simulated HUDF -09E1,-09E2,-12 and ERS datasets, comparing various selection methods and the three $\beta$ measurement methods. 
We present a re-analysis of the \citet{Dunlop2012} $z\approx7$ LBG sample in Section \ref{refitsection}.  
Strategies for analysing the HUDF12 data are presented in Section \ref{hudf12strategies}, wherein we briefly discuss the effect of Lyman Alpha Emitters in Section \ref{LAEsection}. 
In Section \ref{conclusions}, we present our conclusions and outline our plans for future studies.

Where relevant, we assume a cosmology with $\Omega_0=0.3, \Omega_\Lambda=0.7, H_0=70~\rm{km~s}^{-1}\,\rm{Mpc}^{-1}$ and quote magnitudes in the AB system \citep{Oke1965}. For convenience, we use $B_{435}, V_{606}, i_{775}, z_{850}, Y_{098}, Y_{105}, J_{125}, J_{140}$ and $H_{160}$ to refer to the \emph{HST} ACS F435W, F606W, F775W, F850LP and WFC3/IR F098M, F105W, F125W, F140W and F160W filters respectively.

%%%%%%%%%%%%%%%%%%%%%%%%%%%%%%%%%%%%%%%%%%%%%%%%%%%%%%%%%%%%%%%%%%%%%%%%%%%%
%%%%%%%%%%%%%%%%%%%%%%%%%%%%%%%%%%%%%%%%%%%%%%%%%%%%%%%%%%%%%%%%%%%%%%%%%%%%

\section{Methods of determining $\bmath{\beta}$}
\label{measuringbeta}
In this section, we describe three methods for measuring the UV slope of high-redshift galaxies and show that, for perfect photometry, they yield similar results. 
Later in Section 5.3, we explore their relative strengths for estimating $\beta$ in realistic simulated data.

%%%%%%%%%%%%%%%%%%%%%%%%%%%%%%%%%%%%%%%%%%%%%%%%%%%%%%%%%%%%%%%%%%%%%%%%%%%%

\subsection{Single colour ($\bmath{\beta_{J-H}}$)}
The UV spectral index $\beta$ may be approximated from a single colour using no prior assumptions of the underlying spectrum.
Where the colour comprises two filters comfortably redward of the Lyman break it is insensitive to small errors in the photometric redshift. 
Moreover, IGM absorption and any \Lya\ emission present do not contaminate the continuum slope measurement through filters fully redward of 1216~\AA\ in the rest-frame.
With WFC3 photometry of objects at $z>6.5$,
\begin{equation}
\beta_{\rm{colour}}=\beta_{J-H}=4.43(J_{125}-H_{160})-2
\label{eqn:JmHcolour}
\end{equation}
is typically used to estimate $\beta$, where the coefficient 4.43 is found from the filter pivot wavelengths \citep[e.g.][]{Tokunaga2005} using $1/[2.5\log({\lambda_H/\lambda_J})]$, where $\lambda_{J}=12486$ \AA, $\lambda_{H}=15369$ \AA\ when the instrument throughput including the detector response function is included \citep{Dressel2011}.

%%%%%%%%%%%%%%%%%%%%%%%%%%%%%%%%%%%%%%%%%%%%%%%%%%%%%%%%%%%%%%%%%%%%%%%%%%%%

\subsection{Power-law ($\bmath{\beta_{\yjh}}$)}
Objects whose rest-frame UV continuum is present in several filters redward of the Lyman break should in principle have their UV slope better constrained by using all the available information.
In the HUDF09 and ERS at $z\approx6.5$, $Y_{\{098|105\}}, J_{125}$, and $H_{160}$ lie redward of the Lyman break and the additional use of the $Y-J$ colour here should improve the constraint on $\beta$ over the use of only a single $J-H$ colour.
By $z\approx7.5$, the Lyman break diminishes the $Y_{105}$-band flux by almost a half, and a power-law fit should begin to approach a single colour.
However it is not immediately clear whether employing the $Y$-band for galaxies in the redshift space ($6.5\leq z\leq 7.5$) in which the Lyman break is travelling through $Y$ will be beneficial, given the potential for a colour-dependent misplacing of the break within the filter. 
Moreover, high equivalent-width \Lya\ emission lines could bias $\beta$ measurements to significantly bluer values than the intrinsic continuum slope, an effect we investigate in Section \ref{LAEsection}.
In the power-law $\beta$ measurements presented below, the photometric redshift of an object is used to build a grid of SEDs with varying power-law $\beta$ values redward of the Lyman break, and zero flux at $\lambda<1216$~\AA. 
Synthetic photometry of each power-law SED is created, and an object's $\yjh$ photometry is used to select the best-fitting $\beta$ from the grid via a $\chi^{2}$ fit.

%%%%%%%%%%%%%%%%%%%%%%%%%%%%%%%%%%%%%%%%%%%%%%%%%%%%%%%%%%%%%%%%%%%%%%%%%%%%

\subsection{Best-fitting stellar population synthesis model ($\bmath{\beta_{\rm BC03}}$)}
\label{bestfitmodelsection}
To allow a measure of the rest-frame UV continuum slope unaffected by absorption and emission features, \citet*[hereafter C94]{Calzetti1994} defined ten spectral windows in the rest-frame UV avoiding significant spectral features.
While defined for use on continuum spectra, \citet{Finkelstein2012a} advocate the use of the windows on photometric data via SED fitting.
The use of synthetic population synthesis models allows this `pseudo-spectroscopic' measurement to make full use of the available photometry.
Our implementation of this method uses \textsc{fast} \citep{Kriek2009} to perform SED fitting of multi-band photometry, returning both the photometric redshift and the best-fitting \citet[hereafter BC03]{Bruzual2003} population synthesis model SED.
The C94 windows then select the regions of the SED for the power-law fit (a linear fit of $\log f_{\lambda}$ vs. $\log \lambda$). 
The blue limit of the resultant $\beta$ parameter space, $\beta_{\rm{min}}=-3.2$, is governed by the lowest metallicity ($0.05~\rm{Z}_\odot$) and youngest (1~Myr) simple stellar population included in the grid.
 This differs slightly from the approach of \citet{Finkelstein2012a}, who use \textsc{eazy} \citep*{Brammer2008} to obtain the photometric redshift before further SED fitting with BC03 models (or updated variants). 
However, locking the redshifts with \textsc{eazy} prior to fitting the stellar populations with \textsc{fast} shows no appreciable improvement in the recovery of $\beta$ or photometric redshift with respect to the input values.

%%%%%%%%%%%%%%%%%%%%%%%%%%%%%%%%%%%%%%%%%%%%%%%%%%%%%%%%%%%%%%%%%%%%%%%%%%%%

\subsection{Cross-checking the methods}
\label{crosscheck}

\begin{figure}
\includegraphics[width=3.3in]{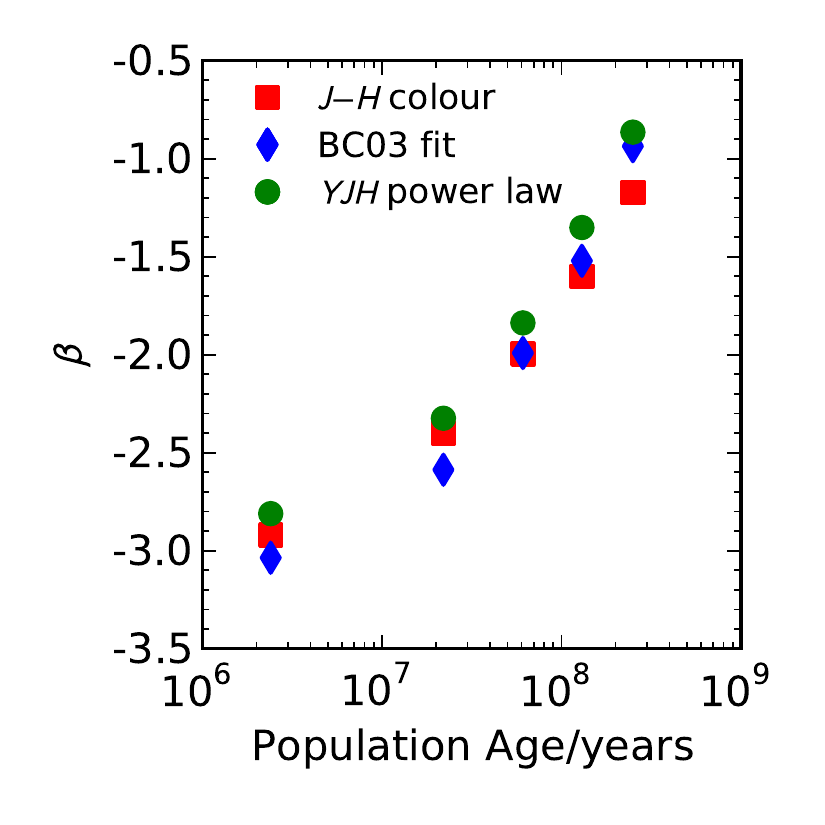}
\caption{The UV slope $\beta$, measured using the three methods discussed in Section 2, for five input SEDs. 
The inputs are various ages of a BC03 population synthesis model at $z=7$ (adopting a \citealt{Chabrier2003} IMF, $0.2~\rm{Z}_{\odot}$ metallicity, and a single burst model). 
Ages of $2.4, 22, 61, 130, 250$~Myr give $\beta\approx$ $-3,-2.5,-2,-1.5,-1$.
Although all three methods agree perfectly for true power-law SED inputs (not shown), it can be seen that small discrepancies ($\Delta\beta\lesssim0.2$) exist for more realistic input SEDs.}
\label{perfectphotom}
\end{figure}

As shown in Fig.\ \ref{perfectphotom}, the three methods generally agree to within $\Delta\beta\lesssim0.2$ when provided with perfect photometry of a BC03 SED.
With such data, we find that the $\beta_{\yjh}$, $\beta_{J-H}$ and $\beta_{\rm BC03}$ methods agree better when the reddest of the ten C94 windows is neglected.
As shown in Fig.\ 1, the reddest C94 window is redward of the $H_{160}$-band at $z\approx7$ and therefore purports to probe a region of the intrinsic SED not covered by the photometry. Thus, any spectral features present in the BC03 models in that region (e.g. the slight jump in flux at 1.8~\micron\ observer-frame in Fig.\ 1) will cause discrepancies between $\beta$ as measured from colours alone and from the best-fitting model. 
We believe this issue may partially explain the offset between $\beta_{J-H}$ and $\beta_{\rm BC03}$ seen in fig.\ 3 of \citet{Finkelstein2012a}.
For this work, we therefore adopt the nine shortest-wavelength C94 windows.

%%%%%%%%%%%%%%%%%%%%%%%%%%%%%%%%%%%%%%%%%%%%%%%%%%%%%%%%%%%%%%%%%%%%%%%%%%%%
%%%%%%%%%%%%%%%%%%%%%%%%%%%%%%%%%%%%%%%%%%%%%%%%%%%%%%%%%%%%%%%%%%%%%%%%%%%%

\section{Simulation methodology}
\label{simulations}
In this section, we present our method for creating mock catalogues of high-redshift galaxies and producing multi-band images of them with realistic noise properties.
The subsequent object recovery, redshift and colour fitting of these simulated galaxies is then described.
The resulting simulations are used thereafter to study the measurement of $\beta$ at $z\approx7$.

%%%%%%%%%%%%%%%%%%%%%%%%%%%%%%%%%%%%%%%%%%%%%%%%%%%%%%%%%%%%%%%%%%%%%%%%%%%%

\subsection{Stellar population choice}
\begin{figure}
\includegraphics[width=3.3in]{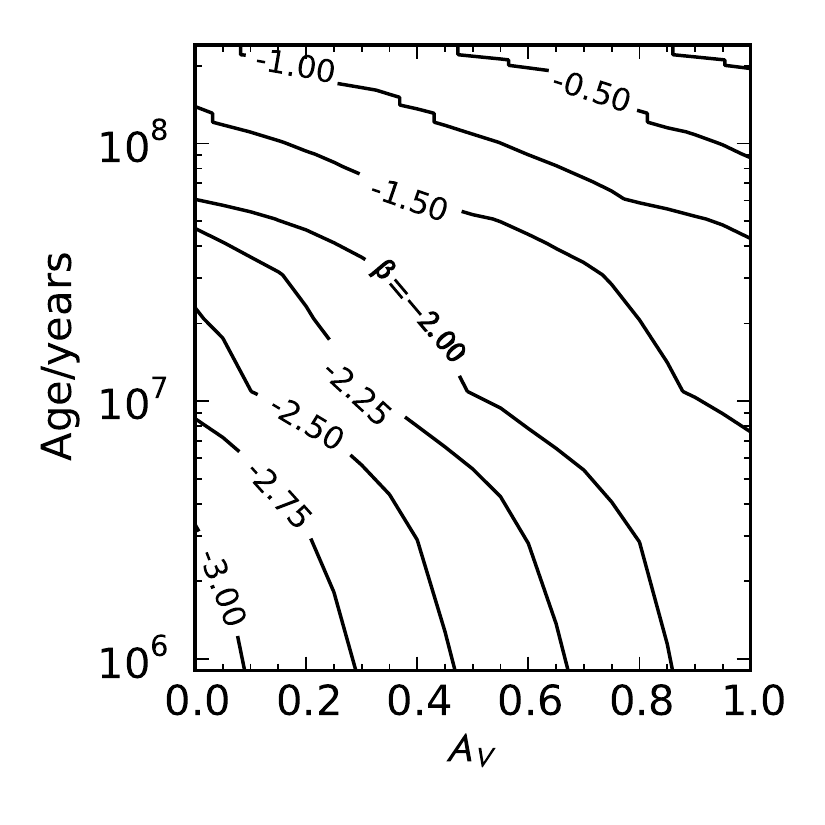}
\caption{Contours showing intrinsic stellar population $\beta$ values in the Age--$A_V$ parameter space. 
$A_V$ parametrizes the dust attenuation, calculated according to the \citet{Calzetti2000} prescription.
The population model is the same BC03 1/5~$\rm{Z}_\odot$ burst used throughout our simulations
(for solar metallicity, add $\approx+0.25$ to each contour's $\beta$ value).
}\label{dust}
\end{figure}

In the simulations used for the remainder of this work, we adopt two model SEDs for simplicity of comparison.
BC03 models of 0.2 Z$_{\odot}$ metallicity with a \citet{Chabrier2003} IMF and ages of 2.4 and 61 Myr are chosen to give SEDs with $\beta_{\rm{in}}\approx-3$ and $\beta_{\rm{in}}\approx-2$ respectively.
These models deviate slightly from perfect power law SEDs, allowing the three $\beta$ measurement methods to yield different results (see Fig. \ref{perfectphotom}). 
However, by using BC03 models in preference to pure power-laws we are better able to realistically represent the true SEDs of high-redshift galaxies.
In our simulations, the input SEDs have not included the reddening due to inter-stellar dust.
Whilst not relevant to the $\beta$ measurement itself -- being only an adjustment of the intrinsic $\beta$ distribution -- it is illustrative to see how low the reddening must be to allow galaxies to be observed with $\beta\approx-3$.
Fig.\ \ref{dust} demonstrates this; for example $\beta=-3$ requires $A_{V}\leq0.1$ and an age under 30~Myr with a low metallicity BC03 model.
However, dust quantities $A_V<1$ (perhaps reasonable at high redshift) cannot redden an SED to $\beta=-1$ before the population is tens or hundreds of Myr old. 
Populations adopting a constant star formation history would additionally require high metallicity to reach $\beta=-1$, as $\beta$ is almost independent of age.
\citet{Wilkins2012} show simulations detailing how the star formation history and metal enrichment history additionally affect the $\beta$ distributions. 

%%%%%%%%%%%%%%%%%%%%%%%%%%%%%%%%%%%%%%%%%%%%%%%%%%%%%%%%%%%%%%%%%%%%%%%%%%%%
\subsection{Simulated image creation}
\label{simplesims}
Our simulation design departs somewhat from that used in recent studies. 
Rather than inserting sources into the real data images, we perform fully synthetic simulations. 
This choice allows a consistent treatment of existing datasets and the anticipated HUDF12 dataset.
In Section 3.3 we verify that this choice does not affect the measured scatter in $\beta$.
Our simulation begins by computing theoretical magnitudes for galaxies of both stellar population models, at a range of redshifts and absolute magnitudes, through the observed filters.
To create sky images, empirical \emph{HST} Point Spread Functions (PSFs) are initially inserted into blank images.
The inserted PSFs are randomly spatially distributed, but pixel-centred, with the relative number density at each redshift slice given by the evolving luminosity function of \citet[eqn. 3]{McLure2009} which reasonably reproduces the observed $z=7$ \citep{McLure2010} and $z=8$ \citep{Bradley2012} functions. 
In the simulations the absolute number density is arbitrarily boosted to allow more robust statistics to be derived.
We mitigate source confusion ($n_{\rm{sources}}\approx0.1$ per aperture) by an arbitrary choice of image size and by neglecting to insert objects at redshifts significantly distant from those of interest. 
For instance, the input catalogue does not feature a $z\sim2$ interloper population (although objects may freely be designated as such `low-$z$ escapees' by the photometric redshift code). 
Redshift $z=$ 5--9 galaxies are included, however, to allow migration of galaxies in-to and out-of the redshift bin of interest, an effect which can have a significant impact on the measured UV slopes.
The luminosity function is integrated down to $M_{\rm UV}(1500\rm{\AA})=-16$, fainter by $\approx2$ mag than the least luminous HUDF objects in the \citet{McLure2011} robust sample of high-redshift galaxies.
Those objects below the detection threshold are useful in providing the simulated image background with some of the non-uniformity seen in real data.

We add artificial noise to these ``perfect'' images, designed to match the noise properties of a given survey.
Table \ref{depthstable} lists the measured $5\sigma$ limiting detection magnitudes for the ERS, HUDF09E1 and HUDF09FULL surveys and estimated limits for the HUDF12.
These depths are computed from the standard deviation of 0.6-arcsec aperture fluxes placed on source-free regions of the images \citep{McLure2011}.
To ensure consistency between these data and our simulations, the simulated noise is designed to match the data's depth when measured in this way, while also accounting for the pixel-to-pixel correlation in the data.
Specifically, we create a noise image for which every pixel is assigned a gaussian random value \begin{equation}
n_{\rm{px}}=\rm{random}\left(\mu=0,~\sigma=\sqrt{\frac{(\sigma_{\rm{aper}}^{\rm{corr}})^2}{\pi r_{\rm{aper}}^2} }~\right),
\end{equation}
where $\sigma_{\rm{aper}}^{\rm{corr}}$ is the $1\sigma$ limiting aperture flux corresponding to $\rm{AB}_{5\sigma}$ -- the limiting detection magnitude of the survey, pre-corrected for pixel-to-pixel correlation by 
\begin{equation}
\sigma_{\rm{aper}}^{\rm{corr}}=\frac{-2.5}{5}\log_{10}\left(\rm{zeropoint}-\left(\rm{AB}_{5\sigma}+\frac{s}{5}\right)\right).
\end{equation}
The noise image is then smoothed with a gaussian filter of standard deviation $s$~px along each axis, which correlates the pixel noise such that the standard deviation of aperture flux in source-free regions yields the desired limiting magnitude, 
yet with global RMS noise (which defines the detection threshold) very comparable to the real data.

\begin{table*}
 \begin{center}
\caption{Limiting magnitudes ($5\,\sigma$ depths) of the datasets considered in this work are shown for \emph{HST} ACS and WFC3 imaging. The ERS and HUDF09E1 depths are measured following the method of \citet{McLure2011} and using their image reductions; HUDF09FULL depths from a consistent treatment of the data released by \citet{Bouwens2012}; approximate depths for the forthcoming HUDF12 WFC3 programme are marked by *. The depths are taken in 0.6-arcsec diameter circular apertures in blank regions of the images as described in \citet{McLure2011}.
HUDF09E1 and HUDF09FULL refer to the first and both epochs of the HUDF09 programme respectively.
$^{\dagger}$While the HUDF12 programme features no further observations in F125W, modest depth improvements are expected from improved reductions of existing data.}\label{depthstable}
\begin{tabular}{ l c c c c c c c c c c }
\hline
 Dataset & $B_{435}$ & $V_{606}$ & $i_{775}$ & $z_{850}$ & $Y_{098}$ & $Y_{105}\phantom{*}$ & $J_{125}$\phantom{*} & $J_{140}$\phantom{*} & $H_{160}$\phantom{*}\\\hline
 ERS & 27.7 & 27.9 & 27.3 & 27.1 & 27.2 & --\phantom{*} & 27.6\phantom{**}& --\phantom{*} & 27.3\phantom{*} \\
 HUDF09E1 & 29.0 & 29.5 & 29.2 & 28.5 & -- & 28.6\phantom{*} & 28.7\phantom{**} & --\phantom{*} & 28.7\phantom{*}\\
 HUDF09FULL & 29.0 & 29.5 & 29.2 & 28.5 & -- & 28.7\phantom{*} & 28.9\phantom{**} & --\phantom{*} & 28.8\phantom{*}\\
 HUDF12 & 29.0 & 29.5 & 29.2 & 28.5 & -- & 29.5* & 29.0*$^{\dagger}$ & 29.0* & 29.0*\\
\hline
\end{tabular}
 \end{center}
\end{table*}

%%%%%%%%%%%%%%%%%%%%%%%%%%%%%%%%%%%%%%%%%%%%%%%%%%%%%%%%%%%%%%%%%%%%%%%%%%%%

\subsection{Object recovery}
For all our simulations, objects are recovered using \textsc{SExtractor} 2.8.6 \citep{Bertin1996} in dual-image mode. 
Objects are selected, unless otherwise noted, from the $J_{125}$ image down to the $1.4\,\sigma$ level (\texttt{DETECT\_THRESH=1.4, THRESH\_TYPE=RELATIVE}) for two adjacent pixels (\texttt{DETECT\_MINAREA=2}).
This method typically selects $\approx10-20\times N_{\rm{input}}$ objects.
Photometry is performed in 10-pixel (0.6-arcsec) diameter apertures for all bands.
The resulting catalogues are cut such that \texttt{MAG\_APER} (detection band) is brighter than the $5\,\sigma\rm{~limit}$, 
retaining approximately 70 per cent of the input objects that were {\it intrinsically} brighter than the $5\,\sigma$ limit.
The fluxes of objects in these catalogues are corrected to total assuming point source aperture corrections for each band.
A $z\approx7$ sample is then obtained by applying a selection function to the catalogue, either a colour-colour cut or a full photometric redshift selection.
In the latter case, we find the redshift and best-fitting stellar population using \textsc{fast} \citep{Kriek2009} with a wide library of BC03 models. For clarity, both input and fitted SEDs contain only simple stellar populations. Age and metallicity are fitted, however, and the models available to \textsc{fast} include those that were used for input.
As \citet{Finkelstein2012a} discuss, the actual choice of models should have little influence on the $\beta$ values, provided a wide range of models are available.
Thus, fully investigating the degeneracies between population parameters is not necessary in order to measure $\beta$ (although see Section 3.4).
Following \citet{Dunlop2012}, we split the sample into \textsc{robust} and \textsc{unclear} categories.
The \textsc{robust} sample contains only galaxies whose primary photometric redshift solution at $6.5\leq z\leq7.5$ is preferred to any secondary solutions by $\Delta\chi^{2}\geq4$.
Galaxies failing to meet this criteria but none the less having a preferred $z\approx7$ solution are denoted \textsc{unclear}. Hereafter, we refer to the combination of \textsc{robust}+\textsc{unclear} as \textsc{all}.

Recovered object candidates are paired to input objects based on their recovered positions. 
Strict position matching ($<2$-pixel radial offset) leaves a sample free of {\it pure} noise spikes ($\lesssim1$ per cent at $5\,\sigma$), yet retains objects where noise spikes, having randomly boosted the flux in individual bands, significantly alter the colours.
Detected objects for which identification of the corresponding input object is ambiguous are similarly dropped ($\sim1$ per cent), although this is minimized by avoiding significant crowding in the simulations.
Objects are only deemed ``ambiguous'' when two or more input objects lie within 2-pixels of a recovered object's position. 
With aperture diameters of 10-pixels, there is still ample ability for faint background objects to contribute to the recovered flux of the detected object.
In summary, our selection criteria are:
\begin{description}
\item $\rm{S/N}(J)\geq5$.
\item $6.5\leq z_{\rm{phot}}(\rm{\textsc{fast}})<7.5$.
\item Sky position is within 2-pixel of a single input object.
\item \textsc{robust}: $\chi^{2}$(secondary $z$) - $\chi^{2}$ (primary $z$) $\geq 4$.
\item \textsc{unclear}:  $0<\chi^{2}$(secondary $z$) - $\chi^{2}$ (primary $z$) $< 4.$
\end{description}
Overall these criteria allow the relevant measurement biases to become manifest while allowing precise tracing of input to recovered parameters.
Having created catalogues with both input and output redshift and photometry parameters, the three $\beta$ measurements for this final sample are made based on the three methods presented above.

%%%%%%%%%%%%%%%%%%%%%%%%%%%%%%%%%%%%%%%%%%%%%%%%%%%%%%%%%%%%%%%%%%%%%%%%%%%%

\subsection{Synthesised noise vs. real noise}

\begin{figure}
\includegraphics[width=3.3in]{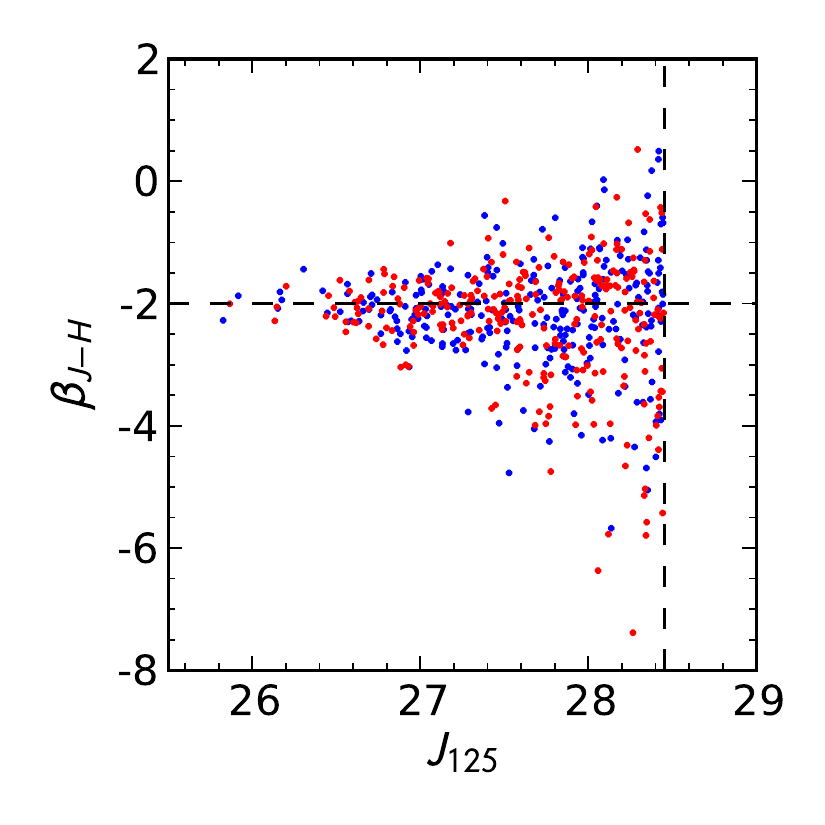}
\caption{
Comparison of the colour-magnitude scatter of simulated $z\approx7$, $\beta_{\rm in}=-2$ galaxies in the HUDF09E1 data using source injection into the real images (blue points) or synthesised noise in blank images (red points). The distribution and scatter of measured UV slopes $\beta_{J-H}$ is very similar in both cases. One third fewer objects are recovered from a source injection simulation (where some inserted objects fall onto existing sources) than in the simulated noise simulations, but for clarity the number of objects in each data series is identical in this figure. In both cases, \textsc{all} (both \textsc{robust} and \textsc{unclear}) $6.5\leq z<7.5$ sources are shown. Dashed lines show the input $\beta$ and the effective $5\,\sigma$ detection limit of the data.
}\label{realfakenoise}
\end{figure}

As discussed above, we have opted to simulate the noise properties of deep \emph{HST} images rather than inject sources directly into the real images. This approach is chosen to allow a consistent treatment of the forthcoming HUDF12 imaging.
We verify that our simulated noise maps yield equivalent results to a source injection scheme by inserting PSFs both into the real HUDF09E1 imaging and into synthetic images with noise designed to match the measured depths of the real data.
Objects are detected and extracted in an identical manner in each case. For the source injection catalogue only, the catalogues are pruned of any objects already present in the unmodified HUDF09E1 images before continuing with the photometric redshift analysis.
Fig.\ \ref{realfakenoise} shows the resulting $\beta\,$--$\,J_{125}$ scatter for each approach. 
Very similar widths in the scatter of $\beta$ are seen at each magnitude, with no significant offset in colour or magnitude between the samples: at $J_{125}=28.0\pm0.25$ both $\sigma({\beta})$ and $\langle\beta\rangle$ differ by $\approx0.1$ 
(simulated noise $\langle\beta\rangle=-2.3, \sigma({\beta})=1.0$; source injection $\langle\beta\rangle=-2.2, \sigma({\beta})=1.0$).
This confirms that the scatter in $\beta$ for faint, low SNR objects is well reproduced by the simulated noise scheme used throughout this work.

%%%%%%%%%%%%%%%%%%%%%%%%%%%%%%%%%%%%%%%%%%%%%%%%%%%%%%%%%%%%%%%%%%%%%%%%%%%%

\subsection{Extended sources vs. point sources}
In the simulations presented here, we have simulated faint $z\approx7$ galaxies with PSFs.
While the faint galaxies our simulations are designed to replicate are very nearly unresolved, we have none the less performed a conservative test of this assumption. We have drawn half-light radii from a gaussian distribution centred on 0.65~kpc with $\sigma$=0.15~kpc, consistent with the size-luminosity results of \citet{Oesch2010a} for $L>0.3 L^{*}$ galaxies at $z\approx7$, and convolved corresponding \textsc{galfit} \citep{Peng2010} models with the PSF.
The measurement of $\beta$ is unaffected if we use these sources in our simulations rather than PSFs.
This is as expected given that our chosen aperture diameter size of 0.6-arcsec corresponds to $\approx5 R_e$ on average.

%%%%%%%%%%%%%%%%%%%%%%%%%%%%%%%%%%%%%%%%%%%%%%%%%%%%%%%%%%%%%%%%%%%%%%%%%%%%
%%%%%%%%%%%%%%%%%%%%%%%%%%%%%%%%%%%%%%%%%%%%%%%%%%%%%%%%%%%%%%%%%%%%%%%%%%%%

\section{Comparison to Dunlop et al. (2012)}
\label{comparedunlopsection}
In contrast to the main simulation approach adopted in this work, \citet{Dunlop2012} injected PSFs into the real HUDF09E1 and ERS images.
Crowding was avoided by inserting sources only within the detection redshift range ($6.5<z_{\rm{in}}<7.5$). 
Furthermore, only objects with $J_{125}^{\rm{input}}<30$ were included -- preventing excess noise being supplied by extra ultra-faint sources.

\begin{figure}
\includegraphics[width=3.3in]{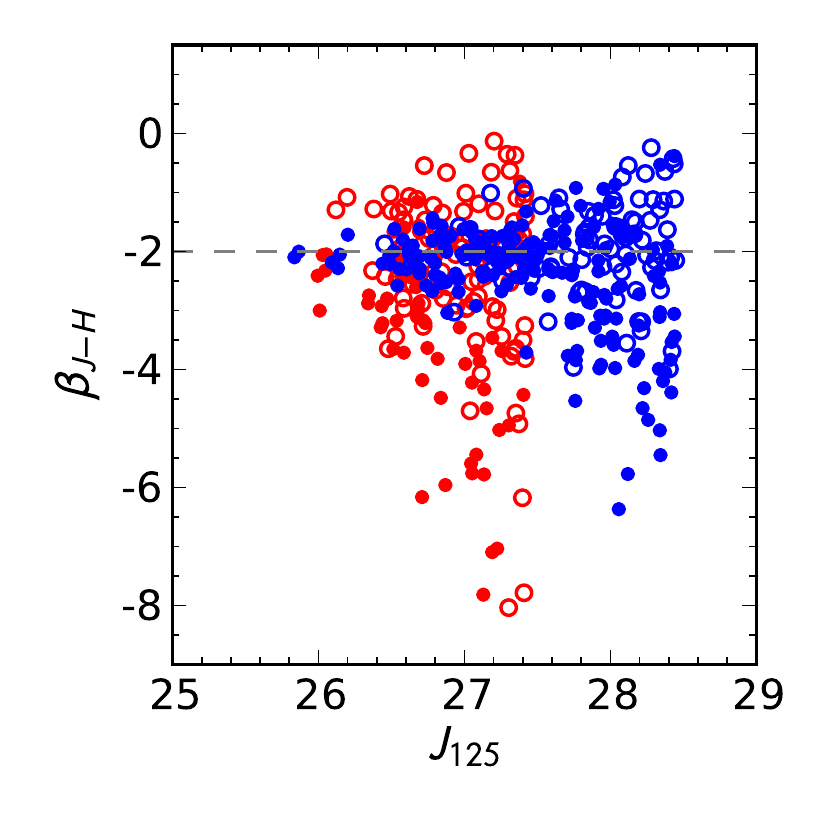}
\caption{Distribution of $\beta$ vs. apparent magnitude from a simulation in which all objects have an intrinsic UV slope of $\beta=-2$. Similar to fig.\ 7 of \citet{Dunlop2012}, all objects have $6.5\leq z_{\rm{in}}, z_{\rm{phot}}<7.5$ and $J_{125}^{\rm{input}}<30$. Red and blue symbols denote the ERS and HUDF09E1 simulations respectively. Open circles denote objects whose photometric redshift is deemed \textsc{unclear}, filled circles are objects with \textsc{robust} photometric redshifts. In contrast to \citet{Dunlop2012}, this plot includes a colour correction of $\Delta(\beta_{J-H})\approx$+0.2 to account for the flux of a point source not enclosed by the photometric aperture in each band.}
\label{dunlopcompare}
\end{figure}

Our new simulations, when limited to the same inputs and selection function, yield results in very good agreement with those of \citet{Dunlop2012}. 
Fig.\ \ref{dunlopcompare} shows the recovered $\beta_{J-H}$ values for a simulated population of faint $\beta_{\rm in }=-2$ objects in the HUDF09E1 and ERS fields, and is remarkably similar to fig.\ 7 of \citet{Dunlop2012}. 
There is a clear offset to blue $\beta$s, which becomes progressively worse for fainter objects. 
In the HUDF simulation, objects in the faintest 1~mag bin average $\langle\beta\rangle=-2.4$. 
This is even  more pronounced in the ERS, where the $J$-band imaging is deeper than the $H$-band imaging and where the $Y_{098}$ filter, which cuts off at a shorter wavelength than the HUDF's $Y_{105}$ filter, is used. 
The bias in the ERS becomes catastrophic for the faintest objects (objects in the faintest 1~mag bin average $\langle\beta\rangle=-2.7$).
The photometric redshifts of even relatively high SNR objects are often deemed \textsc{unclear} when $\beta$ is red, meaning a \textsc{robust} sample excluding such red objects will show a further blueward bias.
Corroborating the work of \citet{Dunlop2012}, we find measuring $\beta$ from a single $J-H$ colour from a sample of $z\approx7$ galaxies yields a large blue  bias for the lowest SNR objects. 

%%%%%%%%%%%%%%%%%%%%%%%%%%%%%%%%%%%%%%%%%%%%%%%%%%%%%%%%%%%%%%%%%%%%%%%%%%%%
%%%%%%%%%%%%%%%%%%%%%%%%%%%%%%%%%%%%%%%%%%%%%%%%%%%%%%%%%%%%%%%%%%%%%%%%%%%%

\section{Discussion of Simulation Results}
\label{simulationresults}
In this section, we use simulations to investigate how both the $z\approx7$ LBG  selection method and the UV slope measurement method affect the measurement of $\beta$.
Simulations, as described above and using $\beta_{\rm{intrinsic}}=-3$ and $-2$, have been constructed of the HUDF -09E1, -09E2, -12 and the ERS datasets.
The depths of these datasets are given in Table \ref{depthstable}.
We use these eight simulations throughout the remainder of this work.

%%%%%%%%%%%%%%%%%%%%%%%%%%%%%%%%%%%%%%%%%%%%%%%%%%%%%%%%%%%%%%%%%%%%%%%%%%%%

\subsection{Comparison of $\bmath z\approx7$ selection functions}
\begin{figure}
\includegraphics[width=3.3in]{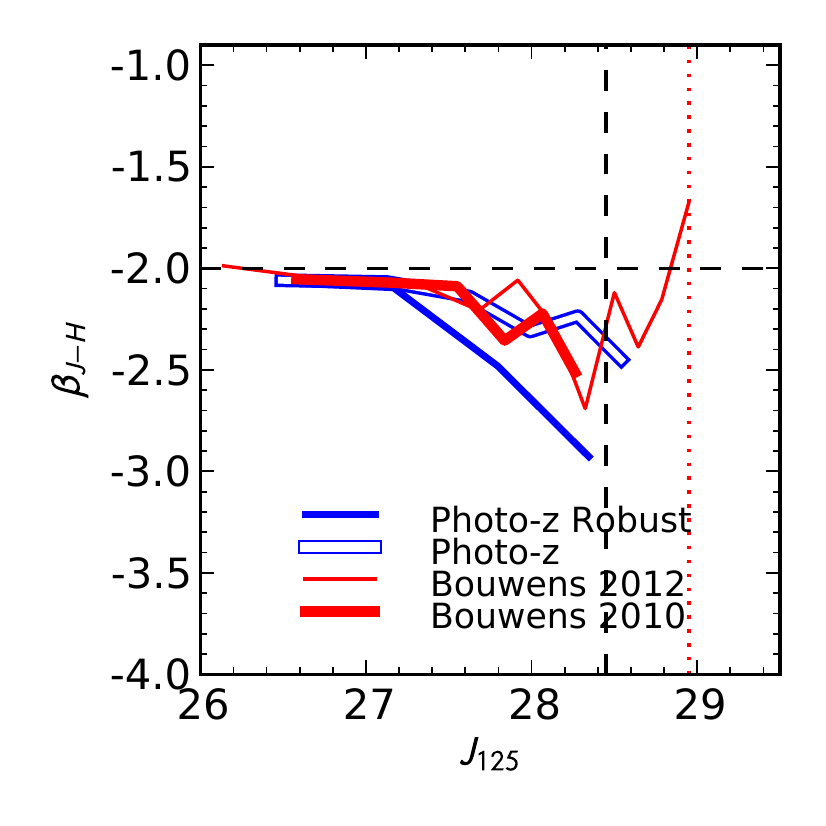}
\caption{Comparison of various selection functions, shown by average UV slopes $\langle\beta_{J-H}\rangle$ from simulated galaxies in the HUDF09E1 in magnitude bins of fixed occupancy. Galaxies have been selected from these images with four selection methods. Thick and thin red lines show objects selected according the the dropout criteria of \citet{Bouwens2010} and \citet{Bouwens2012} respectively. Hollow and solid blue lines show, respectively, objects selected by a full photometric redshift analysis when \textsc{all} $z\approx7$ candidates are included and when only objects with \textsc{robust} photometric redshifts are included. Comparing the two thick lines, we conclude that an inclusive photometric redshift selection and a traditional colour-colour selection suffer similar bias in $\beta$. As expected, a more exclusive photometric redshift selection yields a larger blue bias. The vertical, dashed line shows the effective $5\,\sigma$ flux limit in the $J_{125}-$band. No $J$-band SNR cut is used in the B12 selection method, hence the long tail toward faint magnitudes. In that case, the dotted red line is a guide to the $J_{125}-$band magnitude corresponding to the $5\,\sigma$ limiting magnitude of the stack, assuming $z=7$ and flat-spectrum. These objects, detected from a combined $Y,J,H$ image, only rise above the detection threshold due to noise-spikes boosting the $H$-band flux -- hence their red $\beta_{J-H}$ colours with respect to the input (horizontal, dashed line).}
\label{selectionfunctionplot}
\end{figure}

\label{comparebouwenssection}
Many high-redshift galaxy studies have relied on colour-colour criteria for sample selection rather than using a full SED-fitting photometric redshift code. Here we show an illustrative comparison of those colour-colour criteria employed by \citet{Bouwens2010, Bouwens2012} and of our photometric redshift selection using BC03 template SEDs.

The colour-colour selection criteria described in \citet[hereafter B12]{Bouwens2012} differ from those of \citet[hereafter B10]{Bouwens2010}. 
In both cases, the main selection criteria is a `$z_{850}$-drop': a $z-Y$ colour of $>0.8$ (in B10) or $>0.7$ (in B12). 
Both studies also prohibit the selection of red objects, requiring $Y-J<0.8$. 
B12 use an additional $z-Y$ vs. $Y-J$ colour function, excluding low redshift interlopers that would otherwise have been newly selected following the relaxed $z-Y$ criteria. 
In both studies, various criteria are used to ensure objects with optical detections are excluded.
Crucially, B10 report the use of a $J_{125}\geq5.5\,\sigma$ cut to their catalogue -- a criterion that, as we shall see, is bound produce a bias towards the selection of objects with blue $J-H$ colours. 
This cut is (apparently) abandoned by B12, the faint limit of the catalogue being determined instead by a probability threshold in the detection image (a $\chi^{2}$ image which in this case results in a similar selection to a $Y+J+H$ stack; \citealt*{Szalay1999}).

We have approximately replicated the selection methods of B10 and B12, using our HUDF09E1 $\beta_{\rm in}=-2$ simulation. 
The same simulation is used in both cases to allow a comparison of the selection functions independent of the data variation.
In this case a $\chi^{2}$ detection image, created from the $Y,J,H$ images following the procedure of \citet{Szalay1999}, was used for object detection with aperture photometry performed on individual bands as usual.
Redshift $z\sim7$ galaxies were selected according to the criteria of B10 and B12 independently. 
Two further catalogues were created by selecting  \textsc{all} sources with photometric redshifts $6.5\leq z\leq7.5$, and only \textsc{robust} $6.5\leq z\leq7.5$ sources.
As the colour-colour selections give no precise photometric redshifts, $\beta$ is measured in all cases from the $J-H$ colours.

In Fig. \ref{selectionfunctionplot}, $\langle\beta_{J-H}\rangle$ is shown for each catalogue as a function of $J_{125}$ magnitude. 
We find that the standard photometric redshift selection and the colour-colour selection of B10 are similarly biased toward blue $\beta$ values for faint galaxies. 
A photometric redshift selection is only excessively biased if some additional criteria are used to robustly exclude low redshift interlopers (i.e. $\Delta\chi^{2}\geq4$).

\citet{Bouwens2012} show, in their fig.\ 5, that the B12 selection criteria yield an almost negligible bias in the average UV slope $\langle\beta\rangle$ even for very faint simulated galaxies at $z\approx7$. 
In contrast, Fig.\ \ref{selectionfunctionplot} of this work shows substantial bias in the B12 selected catalogue.
This discrepancy is due to the choice of which observed data are used as a proxy for UV luminosity.
Here we have used $J_{125}$, as this probes rest-frame $M_{1500}$ most closely throughout the $z\approx7$ bin.
As also noted by \citet{Finkelstein2012a}, the clarity of the dependence of $\beta$ on $M_{1500}$ is reduced if one chooses to use $m_{\rm IR}\approx\langle Y,J,H\rangle$ as a proxy for $M_{1500}$.

\begin{figure*}
\includegraphics[width=6in]{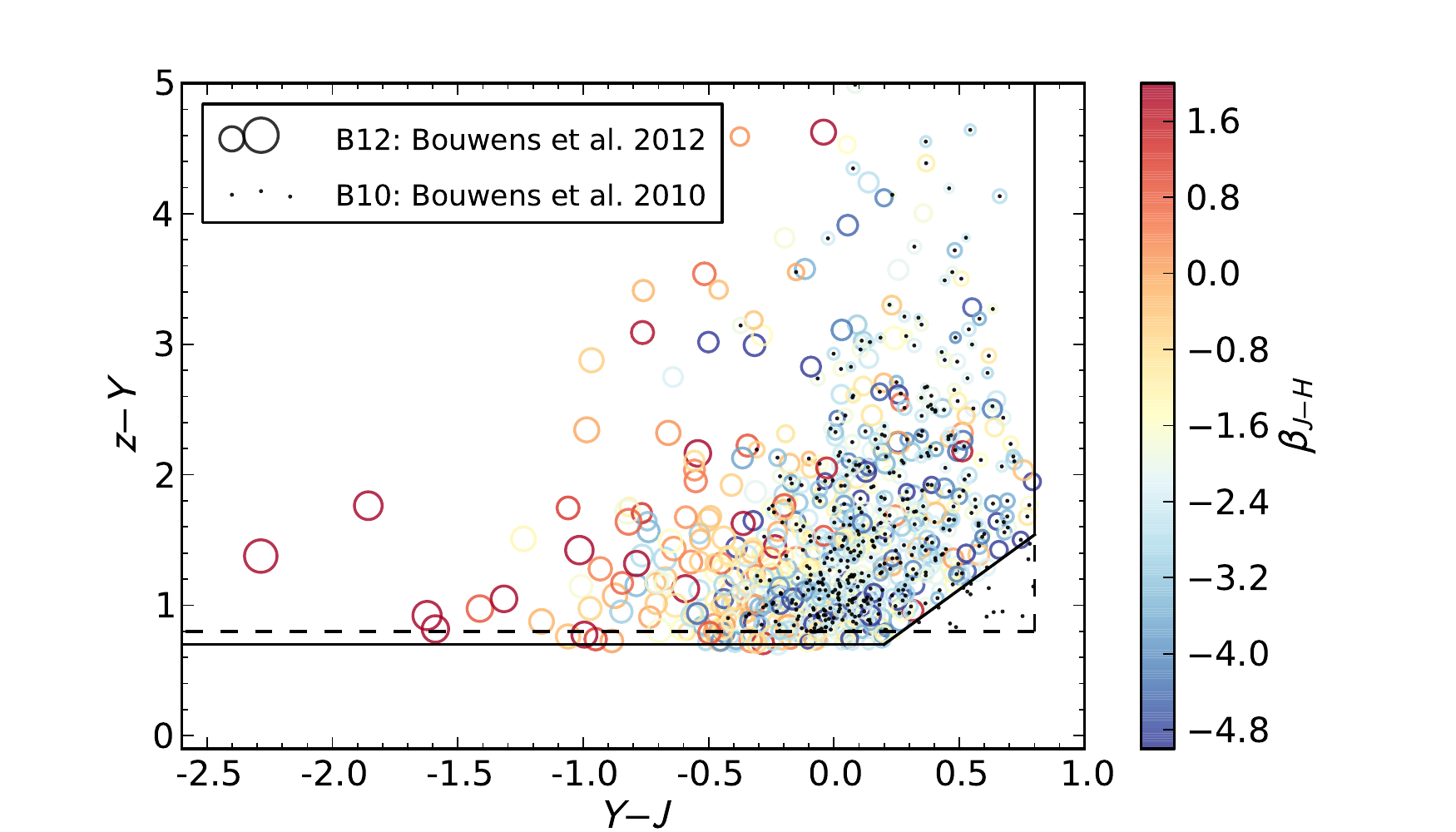}
\caption{
Comparison of the colour-colour selection functions, and resulting samples from simulated images, of B10 \citep{Bouwens2010} and B12 \citep{Bouwens2012}.
Small black dots mark galaxies selected from a HUDF09E1-like simulation, where all galaxies have $\beta_{\rm in}=-2$, using the selection function of B10. The vertical axes shows the Lyman break size ($z\sim7$, $z_{850}$-dropouts). The horizontal axes shows the \Lya-to-UV colour. Coloured circles show a catalogue, selected from the same images, using the B12 selection criteria. Bluer symbols denote steeper UV slopes $\beta$, measured by the $J-H$ colour. Larger symbols denote the galaxies faintest in $J_{125}$ (i.e. largest magnitude). The B12 selection allows galaxies faint in $J_{125}$, with consequently red $J-H$ colours to be selected -- the large, red circles in the lower left of the plot -- which were not selected in B10 due to a $J_{125}\geq5.5\,\sigma$ cut. Objects selected by B10's criteria in the lower right of the plot are treated as contaminants by B12. As can be seen from the colours of nearby objects, many of these would hold blue $J-H$ colours. This combination of changes will clearly allow B12's selection criteria to yield a redder average $\beta$ -- for the same data -- than that of B10.}
\label{bouwenscompare}
\end{figure*}

The difference between the faint ends of the B10 and B12 colour-magnitude relations in Fig. \ref{selectionfunctionplot} is striking.
The removal of an explicit $5.5\,\sigma$ cut in $J_{125}$ by B12 allows many sources with low $J$-band SNR to be included, as seen in Fig. \ref{bouwenscompare}.
In order to be detected in an IR stack, these objects must be flux-boosted in the $H$-band, consequently giving them red $J-H$ colours (the $Y$-band flux is moderately attenuated at $z=7$).
This is clear from Fig. \ref{bouwenscompare}, which shows a comparison of the B10 and B12 selection functions based on the same simulation as Fig. \ref{selectionfunctionplot}.
The addition of faint, red-scattered, sources in the selection function of B12  perhaps accounts for why they report a somewhat redder $\langle\beta\rangle\approx-2.7$ than B10 ($\langle\beta\rangle \approx -3.0$) 
for the faintest galaxies.

In summary, we find that an inclusive photometric redshift selected sample and a $z\sim7$ colour-colour selected sample are similarly biased at the faint end by preferential selection of blue-scattered objects.

%%%%%%%%%%%%%%%%%%%%%%%%%%%%%%%%%%%%%%%%%%%%%%%%%%%%%%%%%%%%%%%%%%%%%%%%%%%%

\subsection{Comparison of $\bmath\beta$ measurement methods}

\begin{figure*}
\includegraphics[width=7in]{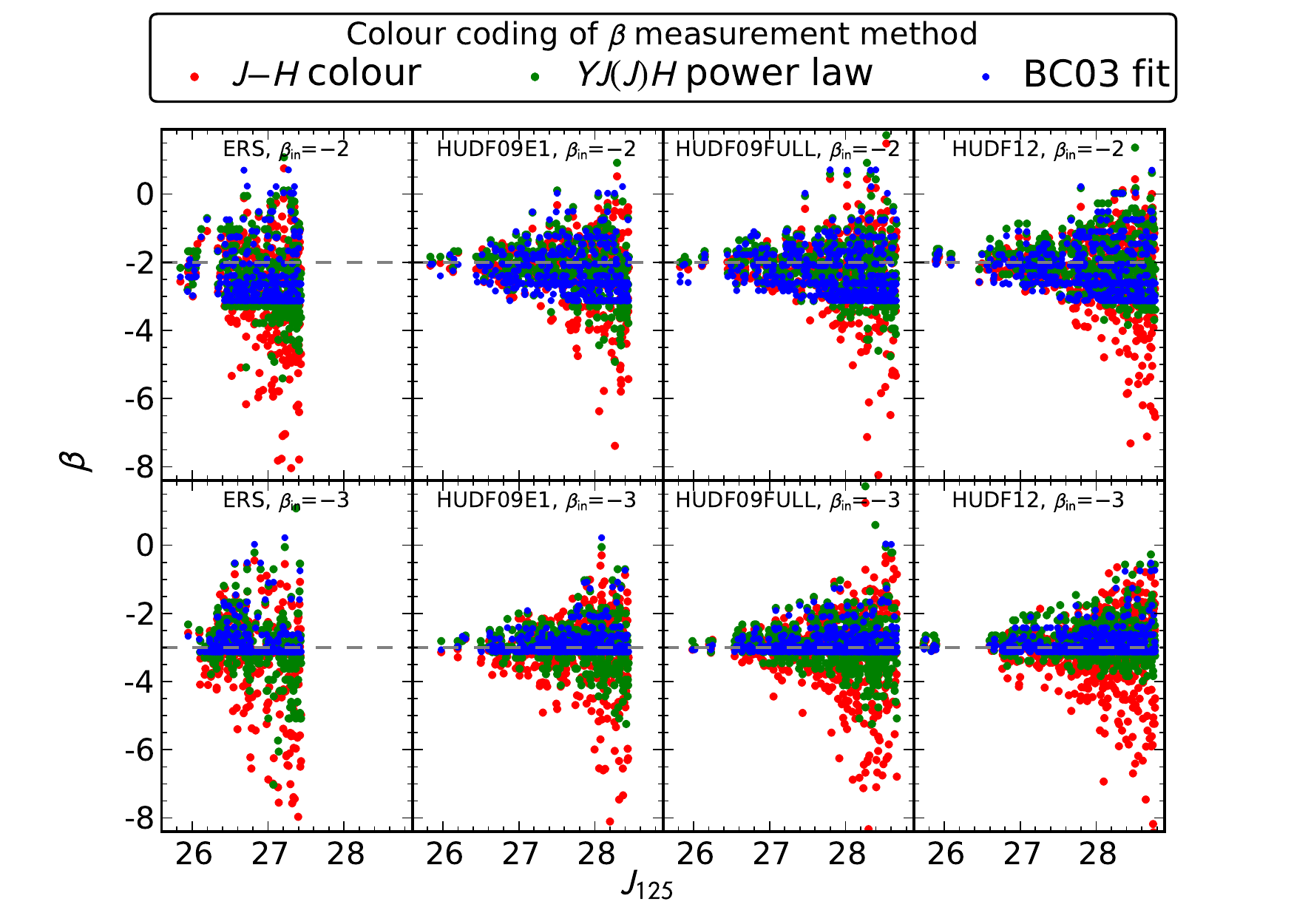}
\caption{Comparison of the three $\beta$ measurement methods, described in Section 2, for simulated objects in the ERS, HUDF09 (epochs 1 and 2) and HUDF12 datasets. For each method (differentiated by colour of symbol), objects' UV slopes are plotted as a function of detection-band magnitude ($J_{125}$, AB mag, corrected to total flux).
Within each simulation, all objects have a single $\beta_{\rm{intrinsic}}$ as shown in the top-right corner of each panel.
\textsc{All} objects with photometric redshift $6.5\leq z\leq7.5$ are included, regardless of the robustness of the redshift fit.
The resulting $\beta$ values as measured from the $J-H$ colour only or a fit to the best-fitting BC03 model's rest-frame UV continuum are shown by red and blue dots respectively.
Green dots show $\beta$ measured using our preferred method: a pure power-law fit to the $\yjh$ photometry ($YJ_{125}J_{140}H$ in HUDF12), attenuated with a Lyman break cutoff to the power-law at $(1+z_{\rm{phot}})\times1216$~\AA.
}\label{3methodsplot}
\end{figure*}

Fig.\ \ref{3methodsplot} shows the recovered UV slopes of simulated objects in the HUDF09E1, HUDF09FULL, HUDF12 and ERS fields as a function of brightness. 
In each simulation, $\beta$ has been recovered using the three methods described above.
Faint objects show extreme scatter in their UV colours, which is maximised when using only a single colour measurement.
The scatter becomes extreme at $\sim0.5$~mag brighter than the $5\,\sigma$ limit in each field.
In the HUDF12, the addition of the $J_{140}$ band primarily benefits the power-law method; although the other methods do benefit indirectly via improved redshift recovery.
In Fig.\ \ref{3methodsplot}, the bias toward faint, blue sources appears similar for the two input $\beta$s (except for when using $\beta_{\rm BC03}$).
If we had considered only \textsc{robust} $z\approx7$ objects, the bias would be more apparent in the $\beta_{\rm in}=-2$ simulation.
Although the selection function is identical for both simulations, where $\beta_{\rm{in}}=-3$ there is a larger colour space available redward of the intrinsic colour. 
Objects can be scattered into this colour space while still being robustly placed at high redshift. 
For example, if a galaxy with $\beta_{\rm in}=-2$ is scattered by $\Delta\beta=+2$ it is liable to be considered a potential low-redshift contaminant and therefore deemed \textsc{unclear}. 
With the same scatter, a galaxy with $\beta_{\rm in}=-3$ will be left with $\beta_{J-H}=-1$ and will likely be kept as a \textsc{robust} high-redshift candidate.

As can be seen in Fig.\ \ref{sedandfiltersplot}, an SED fit at $z=7$ is essentially a fit only to $(z)\yjh$ photometry, with all bluer bands providing non-detections as they lie blueward of the Lyman break. 
It is therefore unsurprising that the $\beta$ distributions as measured from the best-fitting model and from a power-law fit to the $\yjh$ photometry are somewhat similar for $\beta_{\rm{in}}=-2$.
However, objects with \emph{observed} $\beta\lesssim-3$ are unable to have their colours reproduced by the limited parameter space of population synthesis models. 
Thus an apparent tightening of the recovered $\beta$ distribution is seen using the best-fitting model method, by virtue only of an \emph{a priori} assumption of how blue the UV slope may be.
In fact, were the intrinsic colours of faint $z\approx7$ galaxies as blue as $\beta=-3$, a population average of $\langle\beta_{\rm BC03}\rangle$ would not yield this result, but rather a red-biased $\langle\beta\rangle\approx-2.8$ (in the faintest 1 mag bin of our $\beta_{\rm in}=-3$ simulation).
This creates a complicated bias function, since the method returns a \emph{blue} biased $\langle\beta\rangle\approx-2.4$ when $\beta_{\rm in}=-2$ (in the same bin).
The artificial tightening of the scatter is also severe: in the same bin, $\sigma(\beta_{\rm BC03})=0.8$ or 0.5 for $\beta_{\rm in}=-2$ and $-3$ respectively.
For comparison, the faintest 1 mag bin of the $\beta_{\rm in}=-2$ and $-3$ simulations yield $\langle\beta_{\yjh}\rangle=2.3\pm0.9$ and $-3.0\pm0.8$, respectively.
Fundamentally, given that $\langle\beta\rangle$ appears to evolve only mildly with increasing redshift such that the intrinsic $\beta$ distribution likely has an average of $\langle{\beta}\rangle\gtrsim-2.5$, truncating one side of the colour scatter will clearly yield unrepresentative measurements of $\langle\beta\rangle$ and $\sigma(\beta)$.
%%%%%%
\begin{figure*}
\includegraphics[width=7in]{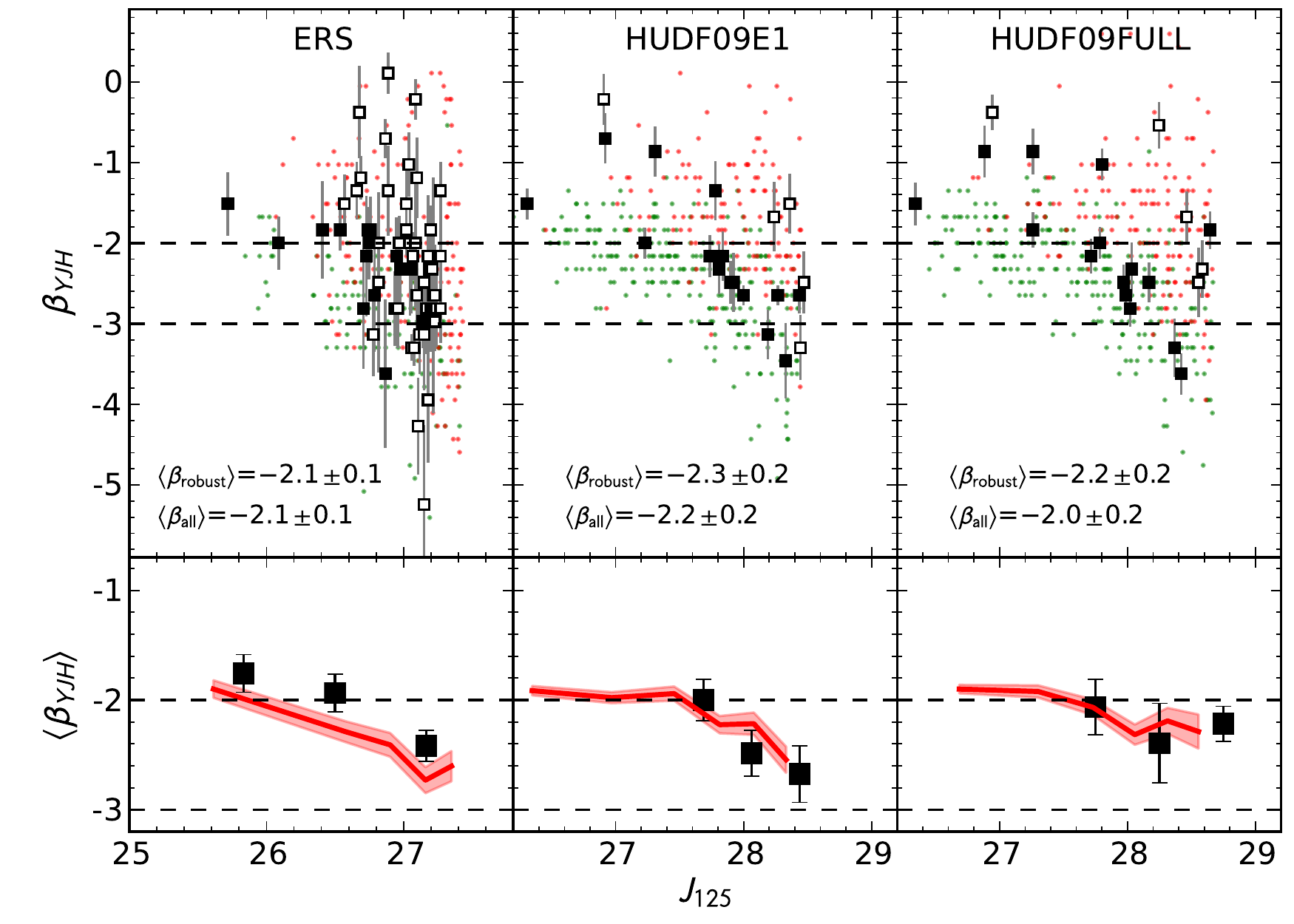}
\caption{HUDF09E1 and ERS $6.5\leq z < 7.5$ galaxies from the \citet{Dunlop2012} sample and our $\beta_{\rm in}=-2$ simulations. In the upper panels, the data are shown as solid (\textsc{robust} photometric redshift) and open (\textsc{unclear} photometric redshift) squares. 
\textsc{Robust} and \textsc{unclear} simulated sources are shown by green and red dots, respectively. 
The right panel shows the sample selected from the HUDF09E1, but with the objects' photometry updated using the HUDF09FULL dataset.
$\beta$ is measured using a Lyman break truncated power-law fit to the available $\yjh$ photometry.
The lower panels show running means, $\langle\beta\rangle\pm\rm{Std.\,Err.}$, for the simulations (red regions) and the faint (well sampled) end of the data binned by magnitude (black squares). In the lower panels, \textsc{all} objects are included.
A clear trend is seen for faint, \textsc{robust} objects to have blue UV slopes; all faint, red objects are assigned \textsc{unclear} photometric redshifts.
At the faint end, this trend is reproduced by our $\beta=-2$ simulation, although the scatter is such that the faintest objects are consistent with a $\beta=-3$ simulation (not shown for clarity, but see Fig.\ \ref{3methodsplot}).
Galaxies in the ERS are slightly redder than a $\beta=-2$ simulation would predict, as are the brightest galaxies in the HUDF.
For each dataset, the inverse-variance weighted mean $\langle\beta\rangle\pm1~$standard error is given both for \textsc{all} sources, and for the \textsc{robust} sub-sample which is consistently slightly bluer.
At the faint end ($J_{125}\geq28$ in the HUDF09, $27$ in the ERS), including \textsc{unclear} sources is more significant. For example their inclusion reddens the faint HUDF09FULL $\langle\beta\rangle$ from $-2.7$ to $-2.3$.
}\label{JBeta}
\end{figure*}
%%%%%%
This is not to discount the use of the best-fitting model method outright: \citet{Finkelstein2012a} have shown strong support for the method in similar source recovery simulations. 
They inserted a population of objects whose distribution of \emph{stellar population parameters} matched those observed in their HUDF sample. 
For the average galaxy in that population, and particularly at $z<7$, \citet{Finkelstein2012a} found that $\beta$ was recovered most successfully from the best-fitting model; a result we reproduce in that the \emph{scatter} is minimal in $\beta_{\rm BC03}$.
\citet{Finkelstein2012a} acknowledge that the method would break down were the parameter space edge to be reached, and the crux of our argument against this method is that the faint, blue, $z\approx7$ galaxies we simulate here exceed that limit (due to the impact of noise).
For galaxies detected with high significance, the method performs well and there is no evidence that their colours are not reproducible by the stellar population synthesis models in an SED fit.
A power-law fit to $\yjh$ photometry, attenuated with a Lyman break as prescribed by the photometric redshift, yields UV slopes closer to their intrinsic values than does the $J-H$ colour, yet without the bias introduced by the assumption that the \emph{observed} colours of low signal-to-noise objects should be reproducible by stellar population models.

The complexity of the bias function for the best-fitting model method (in that it is dependent on the intrinsic $\beta$) is compounded by the reliance on accurate photometric redshifts. 
This reliance is shared by the power-law method.
Where the colours of a galaxy are reproducible by models in the photometric redshift model set, redshift recovery is generally good: $|z_{\rm{phot}}-z_{\rm{in}}|\lesssim0.1$. However, even with perfect HUDF09-like photometry of a galaxy with $\beta_{\rm{in}}=-4$ (pure power law) at $z_{\rm{in}}=7$, a photometric redshift of 6.86 is obtained using BC03 models -- the redshift is underestimated in order to account for the ``excess'' flux in the $Y$-band.
Fortunately, the power-law method is reasonably robust to this: adopting $z=6.86$, a $\yjh$ power-law fit for $\beta$ then yields $\beta\approx-3.8$ -- the value of $\beta$ being tempered slightly toward the colours of the model in the photometric redshift fit.
This bias is clearly still smaller than that seen when $\beta$ is measured directly from the best-fitting model.

%%%%%%%%%%%%%%%%%%%%%%%%%%%%%%%%%%%%%%%%%%%%%%%%%%%%%%%%%%%%%%%%%%%%%%%%%%%%
%%%%%%%%%%%%%%%%%%%%%%%%%%%%%%%%%%%%%%%%%%%%%%%%%%%%%%%%%%%%%%%%%%%%%%%%%%%%

\section{Measurements of $\bmath\beta$ for existing \lowercase{$\bmath{z\approx7}$} galaxy candidates}
\label{refitsection}
Our simulated observations show that, for both the HUDF09 and ERS datasets, a power-law fit to $\yjh$ photometry provides a more reliable measurement of the population's UV slopes than a single $J-H$ colour.
We have therefore re-analysed the photometry of the $z\approx7$ sample of HUDF09 and ERS galaxies, provided by \citet{Dunlop2012}, using a $\yjh$ power-law fit. 
\citet{McLure2011} provide a detailed description of the photometric redshift selection of a similar sample; the \citet{Dunlop2012} sample we use here is more inclusive in that it includes all high-redshift galaxy candidates with both \textsc{unclear} and \textsc{robust} photometric redshifts.
In line with the rest of this work, these catalogues have been pruned of any objects with $J_{125}$-band photometry fainter than the $5\,\sigma$ limit.
In addition, we have updated the photometry of the \citet{Dunlop2012} HUDF09E1 sample using the full-depth HUDF09FULL dataset so that we can compare ERS, HUDF09E1 and HUDF09FULL $z\approx7$ catalogues to our simulations as shown in Fig. \ref{JBeta}.
It is immediately clear that within each dataset there is a trend toward blue $\beta$s at faint magnitudes. 
This is true not only for the \textsc{robust} objects, but for \textsc{all}.
However, this trend is mirrored by the $\beta=-2$ simulation (green and red dots in the figure).
In fact, the faint bin of the HUDF09E1 sample averages $\langle\beta\rangle=-2.6\pm0.2$ which is only marginally bluer than $\langle\beta_{\rm sim}\rangle\approx-2.4$, 
the biased measurement reached by our intrinsically flat-spectrum simulation in the same luminosity bin (see also Fig. \ref{selectionfunctionplot}).
Moreover, as shown in the right-hand panel of Fig. \ref{JBeta}, the higher signal-to-noise delivered by the complete HUDF09FULL dataset yields redder values 
($\langle\beta\rangle=-2.3$ for the faintest luminosity bin) as expected if $\langle \beta \rangle$ is significantly biased by photometric scatter.
Thus, while some of the HUDF's brightest $z\approx7$ galaxies -- which have \textsc{robust} photometric redshifts -- appear redder than $\beta=-2$, there is currently 
no convincing evidence that the faintest objects are significantly bluer than that.

It is of course possible to use a suite of simulations to determine the intrinsic distribution most likely present in the observed galaxy sample. 
In a future paper we intend to investigate the constraints which can be placed on the intrinsic $\beta$ distribution at $z\approx7$, armed with the new HUDF12 dataset.
The new HUDF12 data will also allow us to determine the slope of any colour-magnitude relation present at $z\approx7$. 
There has been gradual convergence toward quantifying this relation at $3<z<6$ \citep[e.g.][]{Finkelstein2012a,Bouwens2012} but, as this work has shown, this remains unclear at $z\gtrsim7$.
The bias toward blue $\beta$ values, which as we have seen is present in a variety of selection methods, is a function of SNR. 
Thus any attempt at constraining a colour-magnitude relation over a wide magnitude-baseline, e.g. using HUDF + ERS + CANDELS data, requires a careful field-by-field approach to the bias correction.

%%%%%%%%%%%%%%%%%%%%%%%%%%%%%%%%%%%%%%%%%%%%%%%%%%%%%%%%%%%%%%%%%%%%%%%%%%%%
%%%%%%%%%%%%%%%%%%%%%%%%%%%%%%%%%%%%%%%%%%%%%%%%%%%%%%%%%%%%%%%%%%%%%%%%%%%%

\section{Strategies for the HUDF12}
\label{hudf12strategies}
The forthcoming HUDF12 programme (GO12498) will provide significantly improved photometry of high-redshift galaxies in three complementary ways. 
First, the depth of the $Y_{105}$ band will be increased to a detection limit of 30~AB mag ($5\,\sigma$, 0.4-arcsec diameter aperture; 29.6~AB in a 0.6-arsec diameter aperture) providing robust photometric redshifts of `$Y$-drop' galaxies at $z\gtrsim8$.
Second, imaging through the additional $J_{140}$ filter will be added reaching a depth equalling that of the current $J_{125}$-band data in the HUDF09FULL (see Table 1). 
Finally, the $H_{160}$ imaging will be increased in depth to match that achieved in $J_{140}$ and $J_{125}$, allowing more secure colour measurements and minimizing 
bias in source selection.
This will allow $Y_{105},J_{125},J_{140},H_{160}$ (hereafter $Y\!JJH$) photometry to be used for fitting the UV SED of galaxies at $z\approx7$, as we have simulated in this work.
A more selection-independent measurement of $\beta$ will also be possible by detecting objects in the $J_{140}$-band, with $J_{125}-H_{160}$ being used as the colour measurement.
Here, we outline the benefits of these two new measurements of $\beta$ at $z\approx7$, each of which will only be possible following the HUDF12 observations.

%%%%%%%%%%%%%%%%%%%%%%%%%%%%%%%%%%%%%%%%%%%%%%%%%%%%%%%%%%%%%%%%%%%%%%%%%%%%

\subsection{$\bmath {J-H}$ colours of $\bmath J_{140}$-selected galaxies}
\label{singlecolourhudf12}
\begin{figure}
\includegraphics[width=3.3in]{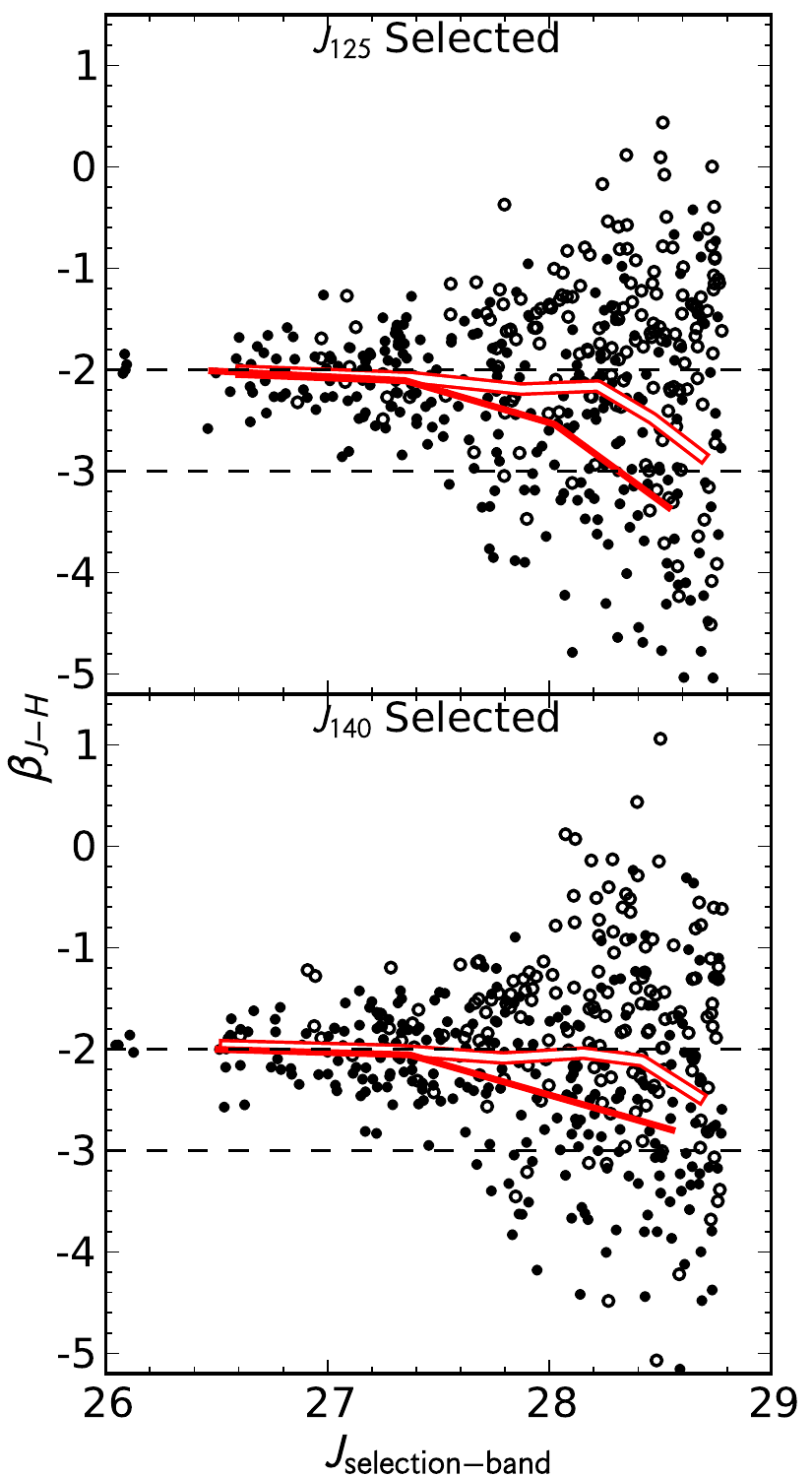}
\caption{
Comparison of the $\beta$ bias in a simulated sample of $z\approx7$ galaxies in the HUDF12, selected in either the $J_{125}$ (upper panel) or $J_{140}$ (lower panel) imaging.
UV slopes, measured via $J_{125}-H_{160}$ colours, are shown for galaxies in our HUDF12 $\beta_{\rm in}=-2$ simulation. Filled and hollow circles mark objects with \textsc{robust} photometric redshifts and \textsc{unclear} objects respectively.
Average UV slope values $\langle\beta\rangle$, in bins of selection band magnitude, are likewise shown by solid (\textsc{robust}) and hollow (\textsc{all=robust+unclear}) lines.
A catalogue produced by selecting objects in the $J_{140}$-band (lower panel) shows a less blue-biased $\langle\beta\rangle$ for faint galaxies than does a $J_{125}$-band selected catalogue (upper panel). This is due to selection band flux boosting in the $J_{125}$-selected catalogue fostering a sub-sample of sources which is very blue in $J-H$ (see Sections \ref{singlecolourhudf12} and \ref{hudf12powerlaw} for discussion).
The hollow line in the lower panel shows that a $J_{140}$-selected photometric redshift catalogue including all objects (\textsc{robust} and \textsc{unclear}) allows an unbiased average UV slope to be measured to low SNR ($<8\,\sigma$). }\label{hudf12singlecolour}
\end{figure}

As discussed in Section 2, a single colour measurement in $J_{125}-H_{160}$ provides a simple estimate of $\beta$ at $z\approx7$.
While we have seen that selecting galaxies (via \textsc{SExtractor}) in the $J_{125}$-band preferentially selects blue galaxies, yielding biased $\langle\beta\rangle$ values, this can be avoided in the HUDF12 by selecting in the $J_{140}$-band.
This method will alleviate some of the `flux-boosting' induced blue bias that is found when $J_{125}$ is used both to detect and determine the colour of $z\approx7$ galaxies.
To quantify the expected benefit of this approach, galaxies from our HUDF12 $\beta_{\rm in}=-2$ simulation have been independently selected in both the $J_{125}$- and $J_{140}$-bands.
Photometric redshift selection of $z\approx7$ galaxies is performed using all bands for both of these catalogues.
In Fig.\ \ref{hudf12singlecolour}, the average UV slope $\langle\beta\rangle$ is shown as a function of selection band magnitude for each catalogue.
Measurement of $\langle\beta\rangle$ for bright galaxies is not affected by the choice of selection band, but within 1~mag of the $5\,\sigma$ image depth a $J_{140}$-selected catalogue clearly provides a less blue-biased measure of $\langle\beta\rangle$ than a $J_{125}$-selected catalogue.
An inclusive $J_{140}$-selected photometric redshift catalogue allows an essentially unbiased measurement of $\langle\beta\rangle$, and $J_{140}$-selection somewhat reduces the bias in a \textsc{robust} photometric redshift catalogue.
Reassuringly, an unbiased measurement of $\langle\beta\rangle$ is possible without resorting to the artificial neglect of certain bands from the photometric redshift analysis as was suggested by \citet{Bouwens2012}.

%%%%%%%%%%%%%%%%%%%%%%%%%%%%%%%%%%%%%%%%%%%%%%%%%%%%%%%%%%%%%%%%%%%%%%%%%%%%

\subsection{Power-law $\bmath\beta$ measurements}
\label{hudf12powerlaw}
\begin{figure}
\includegraphics[width=3.3in]{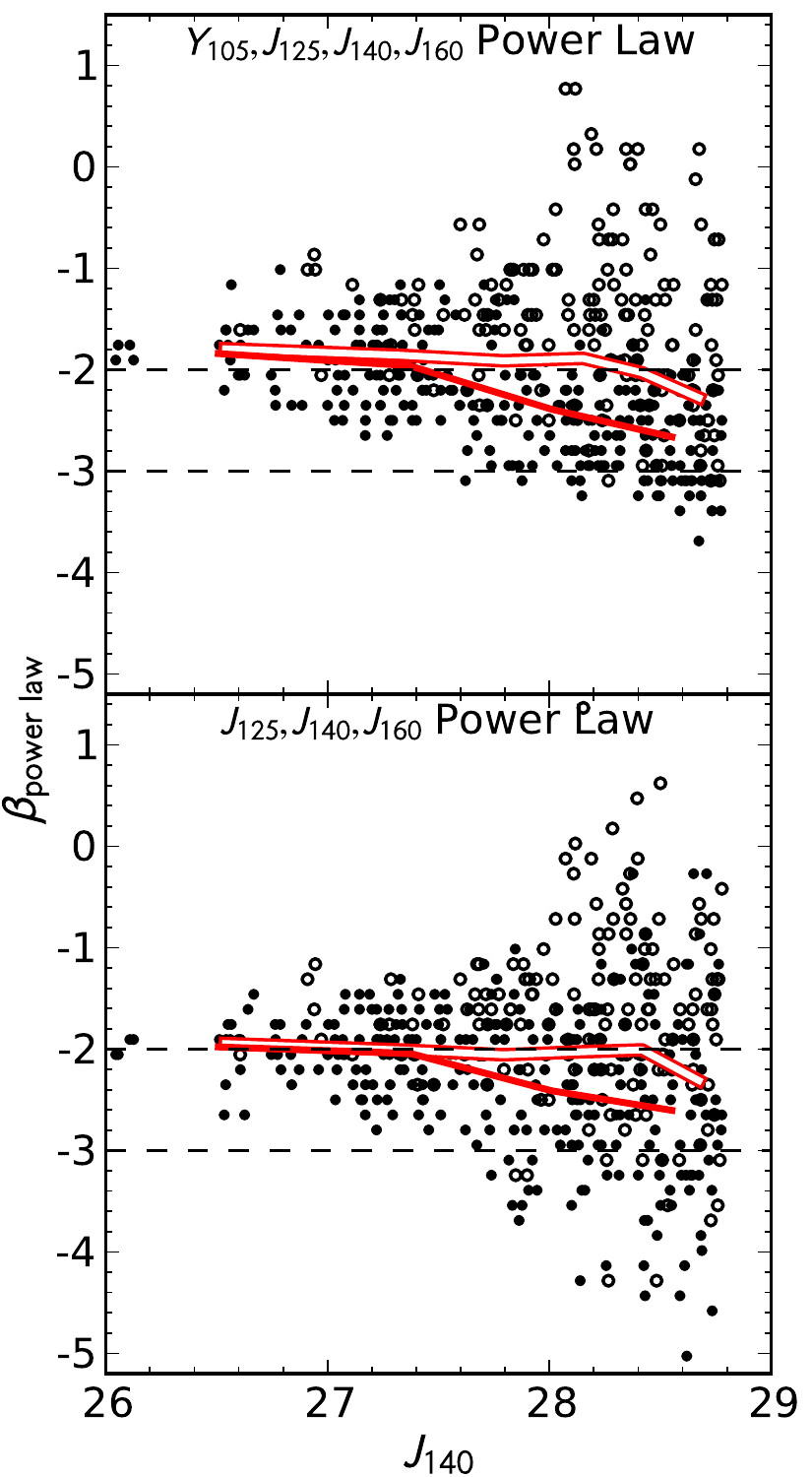}
\caption{
Comparison of the $\beta$ bias in a simulated sample of $z\approx~7$ galaxies in the HUDF12, with $\beta$ measured by a Lyman break truncated power-law fit to $Y_{105},J_{125},J_{140},H_{160}$ (upper panel) or only $J_{125},J_{140},H_{160}$ (lower panel).
Filled and hollow circles mark objects with \textsc{robust} photometric redshifts and \textsc{unclear} objects respectively.
Average UV slope values $\langle\beta\rangle$, in bins of selection band magnitude, are likewise shown by solid (\textsc{robust}) and hollow (\textsc{all=robust+unclear}) lines.
For some objects, the blueward scattering of the $J-H$ colour is tempered by the inclusion of the $Y-J$ colour, although this primarily reduces the width of the scatter and does little to alter $\langle\beta\rangle$. On an average basis, there is therefore little benefit to including the Lyman break affected $Y$-band photometry in the fit.
}\label{YJJHorJJH}
\end{figure}

We have seen that a $J_{140}$-selected catalogue, with $\beta$ measured via the independent $J_{125}-H_{160}$ colour, is less blue biased than a $J_{125}$-selected catalogue.
This is also true when $\beta$ is measured via a power-law fit to $Y\!JJH$, but for more subtle reasons.
The primary cause of bias in $\beta$ is flux boosting of faint objects to just above the detection threshold. 
Galaxies boosted in $J_{125}$ are bound to be measured blue: the colour is always blue relative to both $J_{140}$ and $H_{160}$.\footnote{The $Y$-band photometry carries lower weight, being partially attenuated by the Lyman break.}
However galaxies boosted in $J_{140}$ hold a blue $J_{140}-H_{160}$ colour but a \emph{red} $J_{125}-J_{140}$ colour of similar SNR.
Thus, power-law $\beta$ measurements benefit from a $J_{140}$-selection to a similar degree as the $\beta$ measurements obtained via the $J-H$ colour.
The proposed addition of $J_{140}$ photometry in the HUDF12 also makes possible a multi-band power-law fit to the UV continuum  neglecting the $Y_{105}$-band in which the Lyman break falls.
However with the $Y$-band offering the greatest depth in the HUDF12, it is not immediately obvious whether its exclusion from the $\beta$ measurement will be beneficial or not.
Using the $J_{140}$-selected catalogues described in \ref{singlecolourhudf12}, $\beta$ has been measured, separately, using truncated power-law fits to $JJH$ and $Y\!JJH$ photometry.
From the results shown in Fig.\ \ref{YJJHorJJH}, we can see that the inclusion of the Lyman-break affected $Y$-band photometry does not appreciably reduce the bias in the average UV slope at faint magnitudes. 
While the inclusion of the $Y$-band greatly benefited the measurement of $\beta$ in the HUDF09, it can be excluded in the HUDF12 with only the modest cost of an increase in the scatter of $\beta$ for the faintest galaxies. This is beneficial as a robust measurement of $\langle\beta\rangle$ can then be obtained, via $JJH$ photometry, without relying on the Lyman break affected $Y$-band.

%%%%%%%%%%%%%%%%%%%%%%%%%%%%%%%%%%%%%%%%%%%%%%%%%%%%%%%%%%%%%%%%%%%%%%%%%%%%

\subsection{Lyman-\greektext$\bmath\alpha$ \latintext emitter contamination}
\label{LAEsection}

\begin{figure}
\includegraphics[width=3.3in]{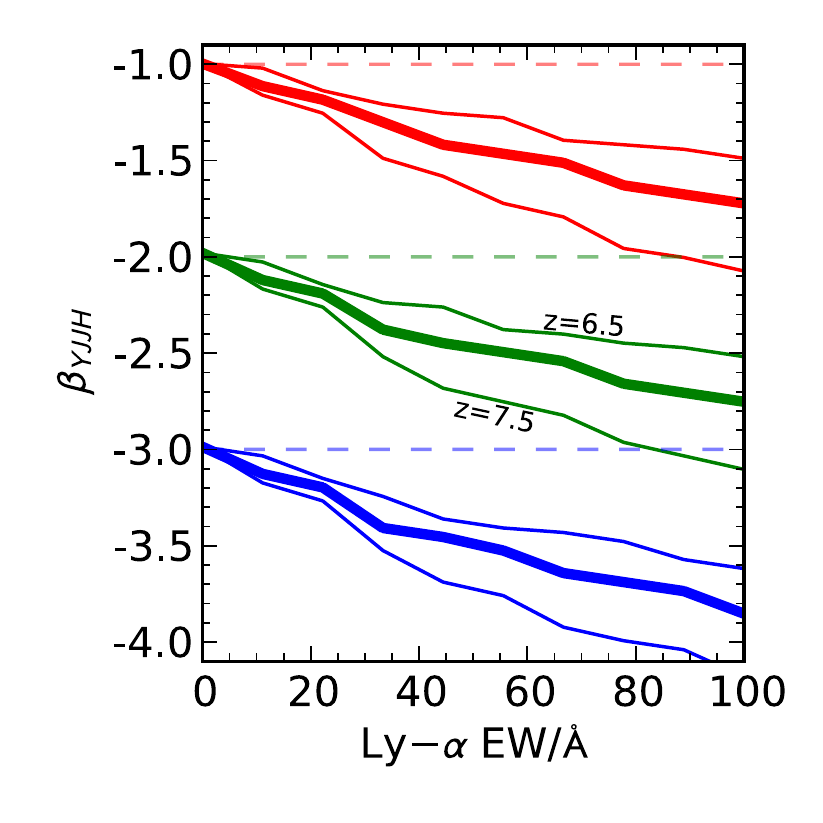}
\caption{The effect of \Lya\ emission of various equivalent-widths on the recovery of the UV slope $\beta$. Perfect power-law spectra, with $\beta=\{-1,-2,-3\}$, were created and the flux at 1216~\AA\ boosted to include an emission line of the specified equivalent-width. Thick red, green and blue lines denote the recovered $\beta$s at $z=7$ for each intrinsic $\beta$ respectively; thin lines (upper/lower) at $z=6.5/7.5$. Here we assume perfect redshift and photometric recovery in the HUDF12's $Y_{105}, J_{125}, J_{140}$ and $H_{160}$ bands.
}\label{laeyjh}
\end{figure}

In the HUDF12 simulations, we have seen that the inclusion of $Y$-band photometry only mildly improves the measurement of $\beta$ in LBGs with no \Lya\ emission.
However, the $Y$-band at $z\approx7$ probes the \Lya\ line and the comparison of $JJH$ to $Y\!JJH$ fits may be a useful check for the presence of \Lya\ emission. 
Furthermore, the existing $\beta_{\yjh}$ measurements for HUDF09 galaxies could potentially be affected by \Lya\ emission; excluding the $Y$-band photometry for these fits would reduce the measurement to only a single $J-H$ colour.

We have performed simple simulations of the effect of \Lya\ emitters (LAEs) on recovered $\beta$ values as follows. 
First, pure power-law spectra were created with $\beta_{\rm{in}}=\{-3,-2,-1\}$, truncated blue-ward of $(1+z)\times1216$~\AA. 
By integrating under the rest-frame UV continuum in the range 1216~\AA\ -- 1216+EW~\AA, the \Lya\ line flux was calculated and added to the flux at 1216~\AA.
Fig.\ \ref{laeyjh} shows the impact of \Lya\ emission of various equivalent-widths on the power-law derived $\beta$ value from $Y_{105}, J_{125}, J_{140}, H_{160}$ photometry (as is present for the HUDF12 -- the effect is marginally stronger in the HUDF09 without $J_{140}$ imaging).
Based on observations out to $z\approx6$, \citet*{Stark2011} predict, in their faint luminosity bin ($26.7<~J_{125}~<~28.2$), a $z=7$ \Lya\ EW distribution peaked in the $25<~\rm{~EW}<~55$~\AA\ bin. 
Galaxies with EW~$>85$~\AA\ represent $\sim5$\% of the population.
Thus we can expect a bias of $\Delta\beta\approx-0.5$, if the LAE fraction does not tail off at $z\approx~7$.
(However, \citet{Bolton2012} have shown that only a small ($\approx$10\%) neutral fraction in the IGM may be sufficient to significantly reduce the transmission of \Lya, and therefore the typical EW of \Lya\ at $z\approx7$.)

This bias is maximised by not floating the redshift here. In practise, the photometric redshift code will select a lower redshift thereby accounting for excess $Y$-band flux by a lower-wavelength Lyman break. 
At $z=7$, 50~\AA\ EW \Lya\ can be countered by misplacing the redshift by $\Delta z\approx-0.3$.
Moreover, the inclusion of \Lya\ emission in photometric redshift fits is commonplace. In principle this allows $z$, $\beta$ (continuum) and the EW of \Lya\ to be correctly determined.

In fact, for our sample of HUDF09 and ERS objects, including the $Y$-band in the measurement of $\beta$ typically returns colours redder than when using only $\beta_{J-H}$.
Thus any \Lya\ present in those galaxies is not boosting the $Y$-band flux to the extent that $\beta_{\yjh}$ is measured with a blue bias.
We can therefore conclude that either no high EW \Lya\ is present, or that the low contribution of \Lya\ is readily countered by an underestimated photometric redshift.
Finally, as we have seen, the addition of $J_{140}$ in the HUDF12 programme will render the $Y$-band photometry unnecessary in fitting $\beta$ at $z\approx7$, alleviating this problem in future analyses of HUDF data.

%%%%%%%%%%%%%%%%%%%%%%%%%%%%%%%%%%%%%%%%%%%%%%%%%%%%%%%%%%%%%%%%%%%%%%%%%%%%
%%%%%%%%%%%%%%%%%%%%%%%%%%%%%%%%%%%%%%%%%%%%%%%%%%%%%%%%%%%%%%%%%%%%%%%%%%%% 

\section{Conclusions}
\label{conclusions}
We have performed object recovery simulations of $z\approx7$ objects in the HUDF and ERS fields, considering how the choice of selection function and $\beta$ measurement method affect the measured average UV slope $\langle\beta\rangle$.

\begin{enumerate}
\item A robust measurement of the UV slope $\beta$ in the ERS and HUDF09 datasets is obtained when fitting a Lyman break truncated power-law SED to $\yjh$ photometry.
In simulations, this method minimizes the scattering of $\beta$ away from the intrinsic value.
The method performs similarly to a method advocated by \citet{Finkelstein2012a}, but avoids the parameter space issues associated with that method -- whereby the scatter in $\beta$ is artificially reduced for ultra-faint, blue, $z\approx7$ objects -- and outperforms the use of a single $J-H$ colour.
\item Our preferred method for measuring $\beta$ relies on a precise photometric redshift measurement, thus can only be made using a full photometric redshift analysis.
As such, and in contrast to claims elsewhere, we have verified that the total bias on the measurement of $\beta$ is similar for a full photometric redshift selection and for a colour-colour selection.
\item In doing so we have highlighted the sensitivity of recovered UV slope measurements to selection function choices.
In particular, a comparison of the colour-colour selection functions of \citet{Bouwens2010} and \citet{Bouwens2012} goes some way to explaining the difference in the average UV slope for faint $z\approx7$ galaxies reported in those studies.
Furthermore, we reiterate that excess bias in a photometric redshift selection function is only seen when optional criteria are added to robustly reject potential low-redshift interlopers.
\item Using our preferred method, new UV slope measurements for a sample of $z\approx7$ galaxy candidates \citep{Dunlop2012} have been made.
The apparent colour-magnitude relation -- whereby the faintest objects appear bluest -- is well reproduced by a simulation in which the intrinsic UV colours of objects are flat-spectrum ($\beta=-2$).
Thus we find even faint $z\approx7$ objects are able to have their colours reproduced by stellar population models of normal star-forming galaxies -- requiring neither extremely young ages nor exotically low metallicities.
\item We have investigated strategies for minimizing the bias in the measurement of $\beta$ from the HUDF12 dataset, finding that using the new $J_{140}$-band imaging for detection, in combination with our preferred $\beta$ measurement method, will yield results with significantly smaller biases than previous estimates.
\end{enumerate}
In a future paper including HUDF12 data, we aim to estimate the intrinsic distribution which underlies the observed UV colour distribution.

\section*{Acknowledgments}
We thank the anonymous referee for helpful comments on our initial manuscript.
ABR acknowledges the support of an STFC studentship.
RJM acknowledges the support of the Royal Society via a University Research Fellowship and the Leverhulme Trust via the award of a Philip Leverhulme research prize. 
JSD acknowledges the support of the Royal Society via a Wolfson Research Merit award, and also the support of the European Research Council via the award of an Advanced Grant. 
This work is based in part on observations made with the NASA/ESA \emph{Hubble Space Telescope}, which is operated by the Association of Universities for Research in Astronomy, Inc., under NASA contract NAS5-26555.

%\begin{thebibliography}{99}
\bibliographystyle{mn2e}                      % The reference style
\bibliography{Rogers2012BetaReferences_abbrv}       % Multiple bib files.

\begin{thebibliography}{42}
\expandafter\ifx\csname natexlab\endcsname\relax\def\natexlab#1{#1}\fi

\bibitem[{Beckwith {et~al}\mbox{.}(2006)Beckwith, Stiavelli, Koekemoer,
  Caldwell, Ferguson, Hook, Lucas, Bergeron, Corbin, Jogee, Panagia, Robberto,
  Royle, Somerville, \& Sosey}]{Beckwith2006}
Beckwith S. V.~W. {et~al.}, 2006, AJ, 132, 1729

\bibitem[{Bertin \& Arnouts(1996)}]{Bertin1996}
Bertin E., Arnouts S., 1996, A\&AS, 117, 393

\bibitem[{Bolton \& Haehnelt(2012)}]{Bolton2012}
Bolton J.~S., Haehnelt M.~G., 2012, MNRAS, submitted (arXiv: 1208.4417)

\bibitem[{Bouwens {et~al}\mbox{.}(2012)Bouwens, Illingworth, Oesch, Franx,
  Labb\'{e}, Trenti, van Dokkum, Carollo, Gonz\'{a}lez, Smit, \&
  Magee}]{Bouwens2012}
Bouwens R.~J. {et~al.}, 2012, ApJ, 754, 83

\bibitem[{Bouwens {et~al}\mbox{.}(2010{\natexlab{a}})Bouwens, Illingworth,
  Oesch, Stiavelli, van Dokkum, Trenti, Magee, Labb\'{e}, Franx, Carollo, \&
  Gonzalez}]{Bouwens2010a}
Bouwens R.~J. {et~al.}, 2010{\natexlab{a}}, ApJ, 709, L133

\bibitem[{Bouwens {et~al}\mbox{.}(2010{\natexlab{b}})Bouwens, Illingworth,
  Oesch, Trenti, Stiavelli, Carollo, Franx, van Dokkum, Labb\'{e}, \&
  Magee}]{Bouwens2010}
Bouwens R.~J. {et~al.}, 2010{\natexlab{b}}, ApJ, 708, L69

\bibitem[{Bradley {et~al}\mbox{.}(2012)Bradley, Trenti, Oesch, Stiavelli, Treu,
  Bouwens, Shull, Holwerda, \& Pirzkal}]{Bradley2012}
Bradley L.~D. {et~al.}, 2012, ApJ, 760, 108

\bibitem[{Brammer {et~al}\mbox{.}(2008)Brammer, van Dokkum, \&
  Coppi}]{Brammer2008}
Brammer G.~B., van Dokkum P.~G., Coppi P., 2008, ApJ, 686, 1503

\bibitem[{Bruzual \& Charlot(2003)}]{Bruzual2003}
Bruzual G., Charlot S., 2003, MNRAS, 344, 1000

\bibitem[{Bunker {et~al}\mbox{.}(2010)Bunker, Wilkins, Ellis, Stark, Lorenzoni,
  Chiu, Lacy, Jarvis, \& Hickey}]{Bunker2010}
Bunker A.~J. {et~al.}, 2010, MNRAS, 409, 855

\bibitem[{Calzetti {et~al}\mbox{.}(2000)Calzetti, Armus, Bohlin, Kinney,
  Koornneef, \& Storchi‐Bergmann}]{Calzetti2000}
Calzetti D., Armus L., Bohlin R.~C., Kinney A.~L., Koornneef J.,
  Storchi‐Bergmann T., 2000, ApJ, 533, 682

\bibitem[{Calzetti {et~al}\mbox{.}(1994)Calzetti, Kinney, \&
  Storchi-Bergmann}]{Calzetti1994}
Calzetti D., Kinney A.~L., Storchi-Bergmann T., 1994, ApJ, 429, 582

\bibitem[{Chabrier(2003)}]{Chabrier2003}
Chabrier G., 2003, PASP, 115, 763

\bibitem[{Dressel(2011)}]{Dressel2011}
Dressel L., 2011, {Wide Field Camera 3 Instrument Handbook, Version 4.0}.
  STScI, Baltimore, USA

\bibitem[{Dunlop {et~al}\mbox{.}(2012)Dunlop, McLure, Robertson, Ellis, Stark,
  Cirasuolo, \& de~Ravel}]{Dunlop2012}
Dunlop J.~S., McLure R.~J., Robertson B.~E., Ellis R.~S., Stark D.~P.,
  Cirasuolo M., de~Ravel L., 2012, MNRAS, 420, 901

\bibitem[{Erb {et~al}\mbox{.}(2010)Erb, Pettini, Shapley, Steidel, Law, \&
  Reddy}]{Erb2010}
Erb D.~K., Pettini M., Shapley A.~E., Steidel C.~C., Law D.~R., Reddy N.~A.,
  2010, ApJ, 719, 1168

\bibitem[{Finkelstein {et~al}\mbox{.}(2010)Finkelstein, Papovich, Giavalisco,
  Reddy, Ferguson, Koekemoer, \& Dickinson}]{Finkelstein2010a}
Finkelstein S.~L., Papovich C., Giavalisco M., Reddy N.~A., Ferguson H.~C.,
  Koekemoer A.~M., Dickinson M., 2010, ApJ, 719, 1250

\bibitem[{Finkelstein {et~al}\mbox{.}(2012{\natexlab{a}})Finkelstein, Papovich,
  Ryan, Pawlik, Dickinson, Ferguson, Finlator, Koekemoer, Giavalisco, Cooray,
  Dunlop, Faber, Grogin, Kocevski, \& Newman}]{Finkelstein2012}
Finkelstein S.~L. {et~al.}, 2012{\natexlab{a}}, ApJ, 758, 93

\bibitem[{Finkelstein {et~al}\mbox{.}(2012{\natexlab{b}})Finkelstein, Papovich,
  Salmon, Finlator, Dickinson, Ferguson, Giavalisco, Koekemoer, Reddy, Bassett,
  Conselice, Dunlop, Faber, Grogin, Hathi, Kocevski, Lai, Lee, McLure,
  Mobasher, \& Newman}]{Finkelstein2012a}
Finkelstein S.~L. {et~al.}, 2012{\natexlab{b}}, ApJ, 756, 164

\bibitem[{Grogin {et~al}\mbox{.}(2011)Grogin, Kocevski, Faber, Ferguson,
  Koekemoer, Riess, Acquaviva, Alexander, Almaini, Ashby, Barden, Bell,
  Bournaud, Brown, Caputi, Casertano, Cassata, Castellano, Challis, Chary,
  Cheung, Cirasuolo, Conselice, Cooray, Croton, Daddi, Dahlen, Dav\'{e},
  de~Mello, Dekel, Dickinson, Dolch, Donley, Dunlop, Dutton, Elbaz, Fazio,
  Filippenko, Finkelstein, Fontana, Gardner, Garnavich, Gawiser, Giavalisco,
  Grazian, Guo, Hathi, H\"{a}ussler, Hopkins, Huang, Huang, Jha, Kartaltepe,
  Kirshner, Koo, Lai, Lee, Li, Lotz, Lucas, Madau, McCarthy, McGrath, McIntosh,
  McLure, Mobasher, Moustakas, Mozena, Nandra, Newman, Niemi, Noeske, Papovich,
  Pentericci, Pope, Primack, Rajan, Ravindranath, Reddy, Renzini, Rix, Robaina,
  Rodney, Rosario, Rosati, Salimbeni, Scarlata, Siana, Simard, Smidt,
  Somerville, Spinrad, Straughn, Strolger, Telford, Teplitz, Trump, van~der
  Wel, Villforth, Wechsler, Weiner, Wiklind, Wild, Wilson, Wuyts, Yan, \&
  Yun}]{Grogin2011}
Grogin N.~A. {et~al.}, 2011, ApJS, 197, 35

\bibitem[{Koekemoer {et~al}\mbox{.}(2011)Koekemoer, Faber, Ferguson, Grogin,
  Kocevski, Koo, Lai, Lotz, Lucas, McGrath, Ogaz, Rajan, Riess, Rodney, \&
  Strolger}]{Koekemoer2011}
Koekemoer A.~M. {et~al.}, 2011, ApJS, 197, 36

\bibitem[{Kriek {et~al}\mbox{.}(2009)Kriek, van Dokkum, Labb\'{e}, Franx,
  Illingworth, Marchesini, \& Quadri}]{Kriek2009}
Kriek M., van Dokkum P.~G., Labb\'{e} I., Franx M., Illingworth G.~D.,
  Marchesini D., Quadri R.~F., 2009, ApJ, 700, 221

\bibitem[{Lorenzoni {et~al}\mbox{.}(2011)Lorenzoni, Bunker, Wilkins, Stanway,
  Jarvis, \& Caruana}]{Lorenzoni2011}
Lorenzoni S., Bunker A.~J., Wilkins S.~M., Stanway E.~R., Jarvis M.~J., Caruana
  J., 2011, MNRAS, 414, 1455

\bibitem[{Madau(1995)}]{Madau1995}
Madau P., 1995, ApJ, 441, 18

\bibitem[{McLure {et~al}\mbox{.}(2009)McLure, Cirasuolo, Dunlop, Foucaud, \&
  Almaini}]{McLure2009}
McLure R.~J., Cirasuolo M., Dunlop J.~S., Foucaud S., Almaini O., 2009, MNRAS,
  395, 2196

\bibitem[{McLure {et~al}\mbox{.}(2010)McLure, Dunlop, Cirasuolo, Koekemoer,
  Sabbi, Stark, Targett, \& Ellis}]{McLure2010}
McLure R.~J., Dunlop J.~S., Cirasuolo M., Koekemoer A.~M., Sabbi E., Stark
  D.~P., Targett T.~A., Ellis R.~S., 2010, MNRAS, 403, 960

\bibitem[{McLure {et~al}\mbox{.}(2011)McLure, Dunlop, de~Ravel, Cirasuolo,
  Ellis, Schenker, Robertson, Koekemoer, Stark, \& Bowler}]{McLure2011}
McLure R.~J. {et~al.}, 2011, MNRAS, 418, 2074

\bibitem[{Meurer {et~al}\mbox{.}(1999)Meurer, Heckman, \&
  Calzetti}]{Meurer1999}
Meurer G.~R., Heckman T.~M., Calzetti D., 1999, ApJ, 521, 64

\bibitem[{Oesch {et~al}\mbox{.}(2010{\natexlab{a}})Oesch, Bouwens, Carollo,
  Illingworth, Trenti, Stiavelli, Magee, Labb\'{e}, \& Franx}]{Oesch2010a}
Oesch P.~A. {et~al.}, 2010{\natexlab{a}}, ApJ, 709, L21

\bibitem[{Oesch {et~al}\mbox{.}(2010{\natexlab{b}})Oesch, Bouwens, Illingworth,
  Carollo, Franx, Labb\'{e}, Magee, Stiavelli, Trenti, \& van
  Dokkum}]{Oesch2010}
Oesch P.~A. {et~al.}, 2010{\natexlab{b}}, ApJ, 709, L16

\bibitem[{Oesch {et~al}\mbox{.}(2012{\natexlab{a}})Oesch, Bouwens, Illingworth,
  Gonzalez, Trenti, van Dokkum, Franx, Labb\'{e}, Carollo, \&
  Magee}]{Oesch2012}
Oesch P.~A. {et~al.}, 2012{\natexlab{a}}, ApJ, 759, 135

\bibitem[{Oesch {et~al}\mbox{.}(2012{\natexlab{b}})Oesch, Bouwens, Illingworth,
  Labb\'{e}, Trenti, Gonzalez, Carollo, Franx, van Dokkum, \&
  Magee}]{Oesch2012b}
Oesch P.~A. {et~al.}, 2012{\natexlab{b}}, ApJ, 745, 110

\bibitem[{Oke(1965)}]{Oke1965}
Oke J.~B., 1965, ARA\&A, 3, 23

\bibitem[{Peng {et~al}\mbox{.}(2010)Peng, Ho, Impey, \& Rix}]{Peng2010}
Peng C.~Y., Ho L.~C., Impey C.~D., Rix H.-W., 2010, AJ, 139, 2097

\bibitem[{Robertson {et~al}\mbox{.}(2010)Robertson, Ellis, Dunlop, McLure, \&
  Stark}]{Robertson2010}
Robertson B.~E., Ellis R.~S., Dunlop J.~S., McLure R.~J., Stark D.~P., 2010,
  Nat, 468, 49

\bibitem[{Stark {et~al}\mbox{.}(2011)Stark, Ellis, \& Ouchi}]{Stark2011}
Stark D.~P., Ellis R.~S., Ouchi M., 2011, ApJ, 728, L2

\bibitem[{Szalay {et~al}\mbox{.}(1999)Szalay, Connolly, \&
  Szokoly}]{Szalay1999}
Szalay A.~S., Connolly A.~J., Szokoly G.~P., 1999, AJ, 117, 68

\bibitem[{Tokunaga \& Vacca(2005)}]{Tokunaga2005}
Tokunaga A.~T., Vacca W.~D., 2005, PASP, 117, 421

\bibitem[{Wilkins {et~al}\mbox{.}(2011)Wilkins, Bunker, Lorenzoni, \&
  Caruana}]{Wilkins2011a}
Wilkins S.~M., Bunker A.~J., Lorenzoni S., Caruana J., 2011, MNRAS, 411, 23

\bibitem[{Wilkins {et~al}\mbox{.}(2012)Wilkins, Gonzalez-Perez, Lacey, \&
  Baugh}]{Wilkins2012}
Wilkins S.~M., Gonzalez-Perez V., Lacey C.~G., Baugh C.~M., 2012, MNRAS, 424,
  1522

\bibitem[{Windhorst {et~al}\mbox{.}(2011)Windhorst, Cohen, Hathi, McCarthy,
  Ryan, Yan, Baldry, Driver, Frogel, Hill, Kelvin, Koekemoer, Mechtley,
  O'Connell, Robotham, Rutkowski, Seibert, Straughn, Tuffs, Balick, Bond,
  Bushouse, Calzetti, Crockett, Disney, Dopita, Hall, Holtzman, Kaviraj,
  Kimble, MacKenty, Mutchler, Paresce, Saha, Silk, Trauger, Walker, Whitmore,
  \& Young}]{Windhorst2011}
Windhorst R.~A. {et~al.}, 2011, ApJS, 193, 27

\bibitem[{Yan {et~al}\mbox{.}(2011)Yan, Finkelstein, Huang, Ryan, Ferguson,
  Koekemoer, Grogin, Dickinson, Newman, Somerville, Dave, Faber, Papovich, Guo,
  Giavalisco, Lee, Reddy, Cooray, Siana, Hathi, Fazio, Ashby, Weiner, Lucas,
  Dekel, Pentericci, \& Conselice}]{Yan2011}
Yan H. {et~al.}, 2011, ApJ, submitted (arXiv: 1112.6406)

\end{thebibliography}
%\include{../library.bib}
%\end{thebibliography}

\bsp

\label{lastpage}

\end{document}